\documentclass[10pt, twocolumn, floatfix, superscriptaddress, aps, prl]{revtex4-2}
\usepackage{amsmath, amssymb, graphicx, xcolor, colortbl, braket}
\usepackage{graphicx}
\usepackage{epsfig}
\usepackage{makecell}
\usepackage{dcolumn}
\usepackage{bm}
\usepackage{amssymb}
\usepackage{amsmath}
\usepackage{ragged2e}
\usepackage{upgreek}
\usepackage{tabularx}
\usepackage{enumitem}  
\usepackage{float}
\usepackage[colorlinks,citecolor=blue,linkcolor=blue,hyperindex]{hyperref}
\usepackage{hyperref}
\hypersetup{
  linkcolor=blue,
  urlcolor=blue
}

\newcommand{\AppA}{\hyperref[app:A]{A}}
\newcommand{\AppB}{\hyperref[app:B]{B}}
\newcommand{\AppC}{\hyperref[app:C]{C}}
\newcommand{\AppD}{\hyperref[app:D]{D}}
\newcommand{\AppE}{\hyperref[app:E]{E}}
\newcommand{\AppF}{\hyperref[app:F]{F}}

\usepackage{lineno}

\begin{document}

\title{Melting temperature shifts from quantum fluctuations in generalized Wigner crystals}
\author{Aman Kumar}
\affiliation{National High Magnetic Field Laboratory, Tallahassee, Florida 32310, USA}
\affiliation{Department of Physics, Florida State University, Tallahassee, Florida 32306, USA}
\author{Sogoud Sherif}
\affiliation{National High Magnetic Field Laboratory, Tallahassee, Florida 32310, USA}
\affiliation{Department of Physics, Florida State University, Tallahassee, Florida 32306, USA}
\author{Veit Elser}
\affiliation{Department of Physics, Cornell University, Ithaca, NY 14853, USA}
\author{Hitesh J. Changlani}
\affiliation{National High Magnetic Field Laboratory, Tallahassee, Florida 32310, USA}
\affiliation{Department of Physics, Florida State University, Tallahassee, Florida 32306, USA}

\begin{abstract} It is generally believed that quantum fluctuations \textit{collaborate} with thermal fluctuations, effectively reducing transition temperatures (e.g. for melting of charge order). We show that this is \textit{not always} the case and that the interplay between quantum and thermal fluctuations can be \textit{competitive}. We find excellent motivation for addressing this thanks to the discovery of correlated insulating ``generalized Wigner crystal" (GWC) states in hetero-bilayer transition metal dichalcogenide (WS$_2$/WSe$_2$) moir\'e systems [Y. Xu, et al., Nature 587, 214–218 (2020)]. We account for the impact of quantum effects on the melting temperature of GWCs, carrying out finite temperature Lanczos calculations on an extended Hubbard model on the triangular lattice (both with a double-gate screened potential, and the nearest neighbor model) for multiple electron densities. We show that quantum effects capture the shift relative to the classical estimates, which in some cases are more than 50 percent off from the experimental values. Then building on these numerical findings, we provide a qualitative picture that clarifies that while quantum melting of GWC (by increasing the bandwidth) naturally softens the ground state order parameter, it \textit{does not} always decrease the melting temperature; \textit{conversely} it can \textit{increase} it.
To do so we employ a finite temperature perturbation theory, treating the kinetic energy perturbatively on top of a classical Wigner crystal.
Our predictions should be observable in future experiments where the bandwidth can be tuned. 
\end{abstract}

\maketitle

\begin{figure*}
    \centering
\includegraphics[width=\linewidth]{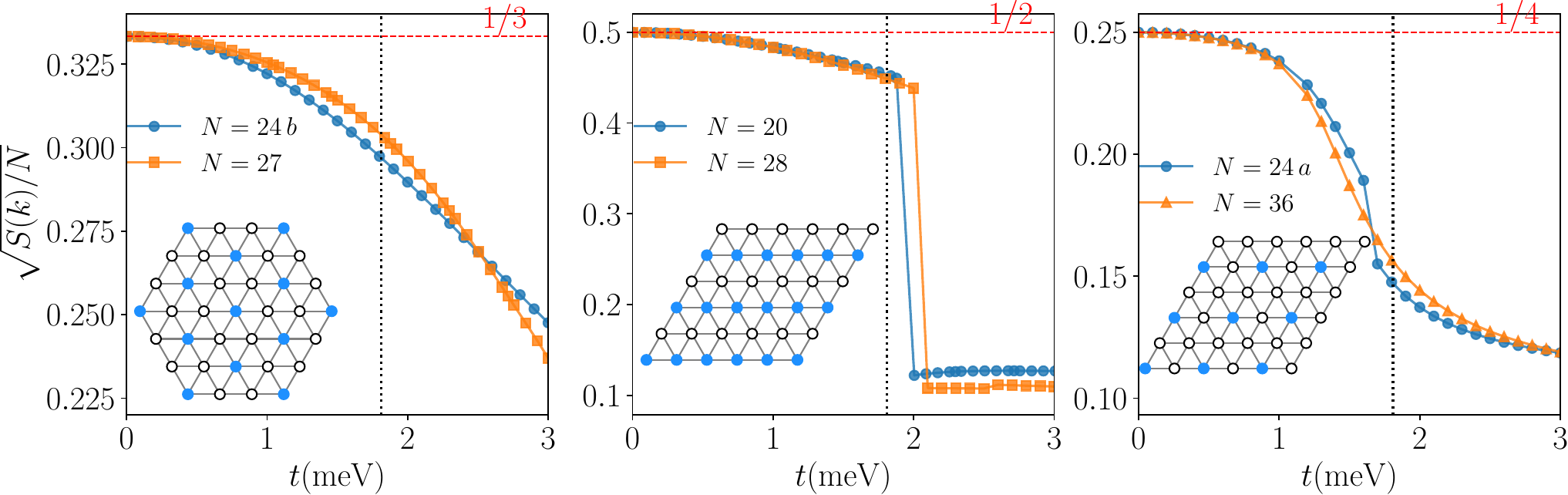}
\caption{Order parameter $\sqrt{S(\mathbf{k})/N}$ for electron fillings $n = 1/3$, $1/2$, and $1/4$, evaluated at the momentum $\mathbf{k}$ where the corresponding classical order parameter is maximal. The dotted red line indicates the classical order parameter value, which equals the particle filling. The results are obtained using a combination of exact diagonalization (ED) and the Lanczos method to access the ground states. 
Dotted lines correspond to $t=1.8$ meV, which is in the ballpark of what has been estimated for the TMD bilayer in previous work~\cite{Motruk_2023}.
Insets show the schematic real-space configurations of the generalized Wigner crystals: a $\sqrt{3}\times\sqrt{3}$ triangular crystal ($n=1/3$), stripe order ($n=1/2$), and a $2\times2$ triangular crystal ($n=1/4$).The GWC for $n=2/3$ (honeycomb) and $n=3/4$ (kagome) are exactly particle-hole dual to their counterparts at $1-n$ (i.e. replace particles by holes and vice versa).}
    \label{fig:schematic_GWC}
\end{figure*}

\textit{Introduction -- } The homogeneous electron gas harbors an intricate phase diagram hosting a Wigner crystal, a periodic arrangement of electrons on a triangular lattice at low density and low temperature, that arises out of effectively strong Coulomb interactions at low density. It was predicted by E. Wigner in 1934~\cite{Wigner_1934}, and subsequently studied theoretically~\cite{Ceperley_Alder_1980, Nagara_1987, Tanatar_Ceperley_1989, Mehta_2013, Zhang_2024}, however, it was only recently imaged in two dimensions in bilayer graphene (in a magnetic field)~\cite{Tsui2024}. In the presence of a triangular moir\'e superlattice-potential, as is realized in hetero bilayer transition metal dichalcogenide (TMD) systems, a class of ``generalized Wigner crystals" (GWCs) has been shown to emerge~\cite{mak_xu2020correlated}. The crystal patterns correspond to triangular, honeycomb, and stripes, as can be seen in Fig.~\ref{fig:schematic_GWC}, and were imaged directly with scanning tunneling microscopy, and indirectly with a combination of optical measurements and classical simulations~\cite{mak_xu2020correlated, feng_li2021imaging}. For a brief overview see Refs.
~\cite{mak_xu2020correlated, feng_li2021imaging, Jin2021, Regan2020, Huang2021, pan2020quantum,Wu_Macdonald,PhysRevB.107.235131,PhysRevB.108.245113, Zhou_review_2025} and references therein.

Considerable effort has been devoted to modeling GWCs, employing a variety of techniques~\cite{Tan_2023,PhysRevB.106.155145,PhysRevB.103.155142, PhysRevLett.132.076503, PhysRevB.103.125146, Matty2022, PhysRevB.103.L241110,PhysRevB.107.235131, Padhi_2021,arxiv.2409.11202}. 
A minimal description is the extended Hubbard model on the (moire) triangular lattice,
\begin{equation}
 H = - \sum_{i, j \sigma} t_{ij} c^{\dagger}_{i,\sigma} c_{j,\sigma}  + U \sum_{i} n_{i,\uparrow} n_{i,\downarrow} + \sum_{i<j} V_{ij} n_i n_j
 \label{eq:model_ham}
\end{equation}
where the symbols have their usual meanings (see Ref.~\cite{kumar_wigner_2025} and references therein) and $n = N_P/N$ is the density, 
where $N_P$ is the number of particles and $N$ is the total number of sites.
We consider the case of the moir\'e bilayer placed symmetrically between two screening gates where the effective screened long-range (LR) potential between charges is given by,
\begin{equation}
V_{ij} \equiv V(r=|\vec{r}_i-\vec{r}_j|) =  \frac{e^2}{4\pi \epsilon \epsilon_0 a }\sum^{k=\infty}_{k=-\infty} \frac{(-1)^k}{\sqrt{\Big(\frac{kd}{a} \Big)^2 + \Big( \frac{|\vec{r}|}{a} \Big) ^2}} 
\end{equation}
where $a$ is the moire lattice constant, and $d$ is the separation between gates. To treat this LR potential, we use the method of images as in Ref.~\cite{mak_xu2020correlated, kumar_wigner_2025}. 
We also consider the nearest neighbor (NN) repulsion model and show that many of the conclusions also apply to this model at finite temperature. 

Both moir\'e and cold-atom systems allow for tunable electron densities, providing access to both metallic and correlated insulating phases. In the moire case, the bandwidth (hopping) can be controlled with an out-of-plane (displacement) electric field~\cite{tang_natcomm_2022}. This enhances quantum effects, softening the order parameter while also altering the transition temperature, \textit{typically} reducing it. Our work here shows this is not always the case, and our results serve as a prediction for possible future experiments. (The original experiments that reported GWCs did not vary the bandwidth~\cite{mak_xu2020correlated}, but this is feasible now.)

\begin{table}
\centering
\begin{tabular}{lccccc}
\hline
 & $n = 1/7$ & $n = 1/4$ & $n = 1/3$ & $n = 2/5$ & $n = 1/2$ \\[5 pt]
\hline
$T_c$ (model)      & 0.0194 & 0.0318 & 0.0824 & 0.0348 & 0.0298 \\[5 pt]
$T_c$ (K, model)   & 8.72   & 14.31  & 37.00  & 15.62  & 13.40 \\[5 pt]
$T_c$ (K, exp)     & $10 \pm 2$ & $32 \pm 2$ & $37 \pm 1$ & $18 \pm 1$ & $29 \pm 2$ \\[5 pt]
$\Delta T_c (K)$   & 1.28   & 17.69  & NA  & 2.38  & 15.66 \\[5 pt]
$\frac{\Delta T_c}{T_c(exp)} \%$ & 13 &  55   & NA & 13 & 54 \\[5 pt]
\hline
\end{tabular}
\caption{Comparison between the classical model and experimental transition temperatures $T_c$ for various densities $n$, using a dielectric constant fixed to 
$\epsilon = 4.67$ by matching the $n = 1/3$ (classical) transition temperature. 
The first row lists the transition temperature in units of $e^2/(4\pi \epsilon \epsilon_0 a)$
obtained from the model,
while the second row shows the corresponding value in Kelvin.}
\label{tab:Tc_eps4p69}
\end{table}

In fact Ref.~\cite{mak_xu2020correlated} modeled classical melting, i.e. $t=0$, and set $U \rightarrow \infty$. 
While $t=0$ is questionable, $U=\infty$  
is reasonable, given the low densities realized in the experiment and the fact that $U/t$ is large. This choice permitted large scale classical Monte Carlo to compute the transition temperature $T_c$, and using it the only unknown parameter, $\epsilon$, was fit to match the recorded temperature for $n=1/3$. (We have recalculated the specific heat and re-fit our data to the experiment to determine $\epsilon$, we report these results in Appendix \AppA). Building on this work, Ref.~\cite{kumar_wigner_2025} considered the case of $n=1/3$ and $n=2/3$ and incorporated quantum effects adding a nearest-neighbor hopping (and a finite but large $U/t$) and estimated that the difference in $T_c$ between the two was in the ballpark of that seen in the experiment. This work also revealed why the long-range nature of the potential was important, and why its effects can be mapped to the widely used NN model for $n=1/3$~\cite{zhou2023quantum}, and why such a mapping may be potentially inadequate for other densities. 

The importance of non-zero $t$ can be seen in Fig.~\ref{fig:schematic_GWC} where we consider the LR model with variable nearest-neighbor $t$ (whose details we elaborate on later in the text). We plot the order parameter (as defined in the caption) in the quantum ground state $|\psi_{GS} \rangle$, which is proportional to the charge structure factor at the ${\bf k}$ point at which it is maximal,
\begin{equation}
S({\bf k}) \equiv \frac{1}{N} \sum_{i,j} e^{i\bf{k \cdot (r_j-r_i)}} \langle \psi_{GS} | n_i n_j | \psi_{GS} \rangle
\end{equation}
Results for representative system sizes are shown and the calculations have been performed with either full diagonalization or Lanczos. The contributions from exactly degenerate ground states have been averaged over, where applicable. The sharp jumps (discontinuities), given the small system sizes should be considered consistent with the existence of a first order quantum phase transition. Such discontinuities arise from the change in momentum sector associated with the ground state leading to energy level crossings as a function of a quantum parameter (here $t$) even on a finite system size. In situations where the momentum sector is unchanged,
such crossings do not exist on finite system sizes, they become avoided crossings due to level repulsion, thus the discontinuities in the order parameter are absent. The dotted line corresponds to $t=1.8$ meV, which has been reported previously in reference to the TMD bilayer~\cite{Motruk_2023}. 

To the best of our knowledge, no previous study has addressed the issue of the considerable difference between the transition temperature of the (classical) theory and experiment for various other fillings, as Table~\ref{tab:Tc_eps4p69} shows. The key ingredient missing is a treatment of quantum effects on melting. While the resolution of the  quantitative discrepancy is important, a pressing physical question arises: 
can quantum fluctuations
of the ground state make the GWC  
ordered state \textit{more} stable to finite temperature fluctuations? We find that it has a nuanced answer that depends strongly on the electron density.

To address this, we systematically simulate finite system sizes with quantum mechanical calculations at zero and finite temperature using two numerical methods. When feasible, we use exact (full) diagonalization using standard quantum statistical mechanics to compute expectation values, 
\begin{equation}
 \langle \hat{O} \rangle = \frac{\textrm{Tr} \Big( e^{-\beta H} \hat{O} \Big)}{\textrm{Tr} \Big( e^{-\beta H} \Big)} = \frac{\sum_i e^{-\beta E_i} \langle \psi_i | \hat{O} | \psi_i \rangle}{\sum_i e^{-\beta E_i}}
\end{equation}
where $\beta=1/T$ is the inverse temperature, $E_i$ refer to the eigenenergies of the $i^{th}$ state and $|\psi_i \rangle$ is the corresponding eigenstate. For systems with larger Hilbert spaces, we use the finite temperature Lanczos method (FTLM)~\cite{ Jaklic_Prelovsek, Prelovsek_2013} that involves approximating the exact sum (trace) by a statistical one
--- results from independent Lanczos runs starting from random vectors are averaged in this procedure. [For work on related systems that employed FTLM, see Refs.~\cite{kumar_wigner_2025, lee2023triangular}. For details of the parameters used in the FTLM see Appendix \AppB.
Many of the finite size clusters have been employed in previous work~\cite{kumar_wigner_2025} or are parallelogram shaped. In Appendix \AppC~we show the triangular clusters that are not typically used, but which have been employed in this work.]

For the case of $n=1/3$ and $n=2/3$, Ref.~\cite{kumar_wigner_2025} provided some evidence that the charge melting transition temperature is weakly affected by the introduction of spin. This is because $U$ is a very large scale in
this system and the charge ordering and magnetism are essentially (but not completely) disentangled in the insulating phase. While we can not rule out the possibly significant role of spin near the GWC-Fermi liquid quantum phase transition, we choose to work with the spin polarized sector throughout this study. This allows us to access system sizes that would not be accessible otherwise within the limitation of our numerical approach.~\footnote{A more detailed analysis of the physics associated with the ground state phase transitions, with spin incorporated, will be addressed elsewhere.}

\begin{figure*}
    \centering
    \includegraphics[width=\textwidth]{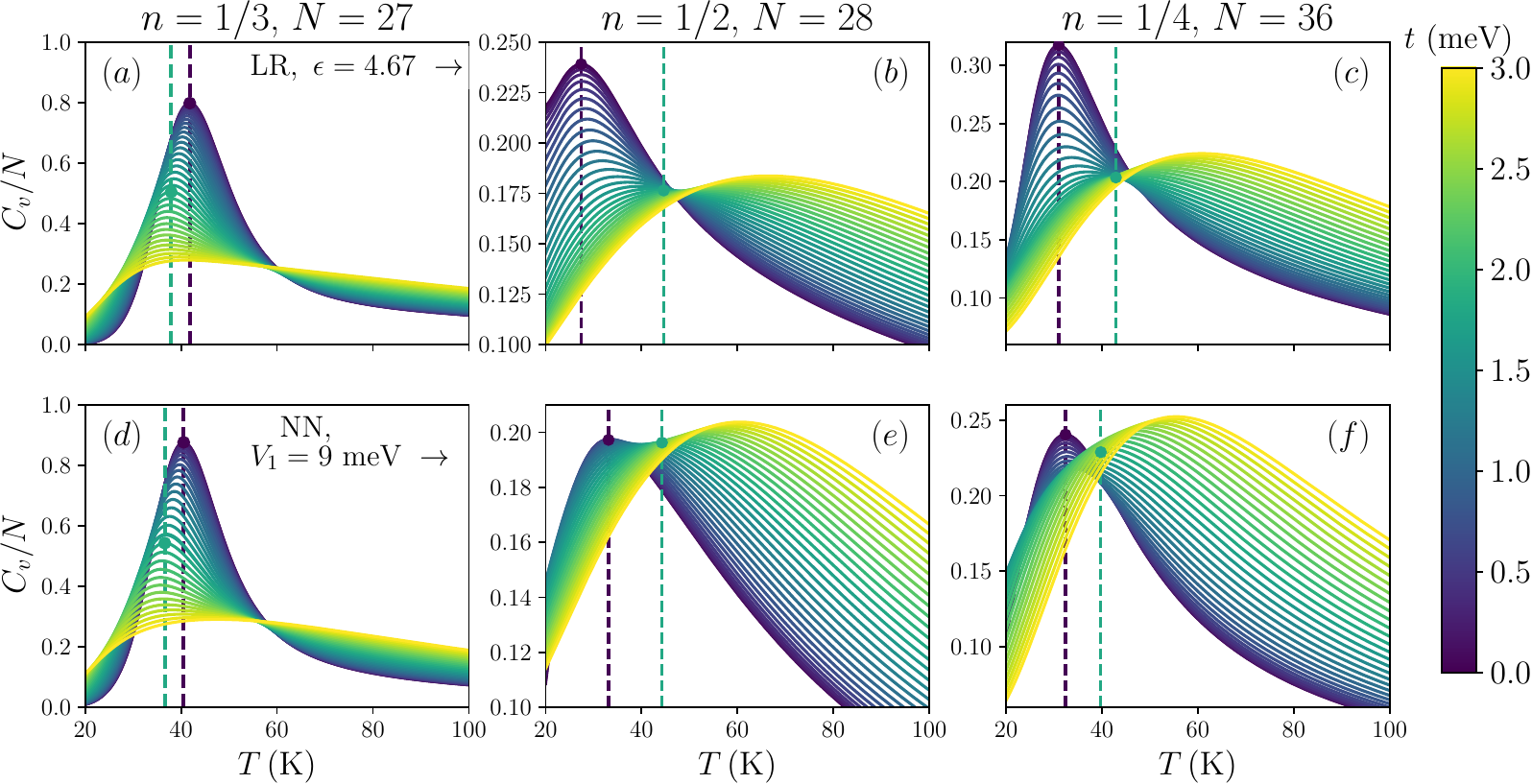}
    \caption{
    Temperature dependence of the specific heat (heat capacity per site in arbitrary units), $C_v/N$, for filling fractions
    $n=1/3$, $1/2$, and $1/4$ (left to right) for a range of $t$ varying from 0 meV (purple) to $3$ meV (yellow) with the experimentally relevant value of $t=1.8$ meV in green. Dashed lines correspond to the $T_c$ for the $t=0$ and $t=1.8$ meV cases. Fully polarized electrons are considered. The top panels [(a-c)] correspond to the (screened) long range model with $\epsilon = 4.67$, $a=7.98$ nm and $d/a = 10$. 
    The bottom panels [(d-f)] correspond to the nearest neighbor (NN) model with $V_1 = 9$ meV. 
    }
    \label{fig:specific_heat}
\end{figure*}

\begin{figure*}[]
    \centering
    \includegraphics[width=\textwidth]{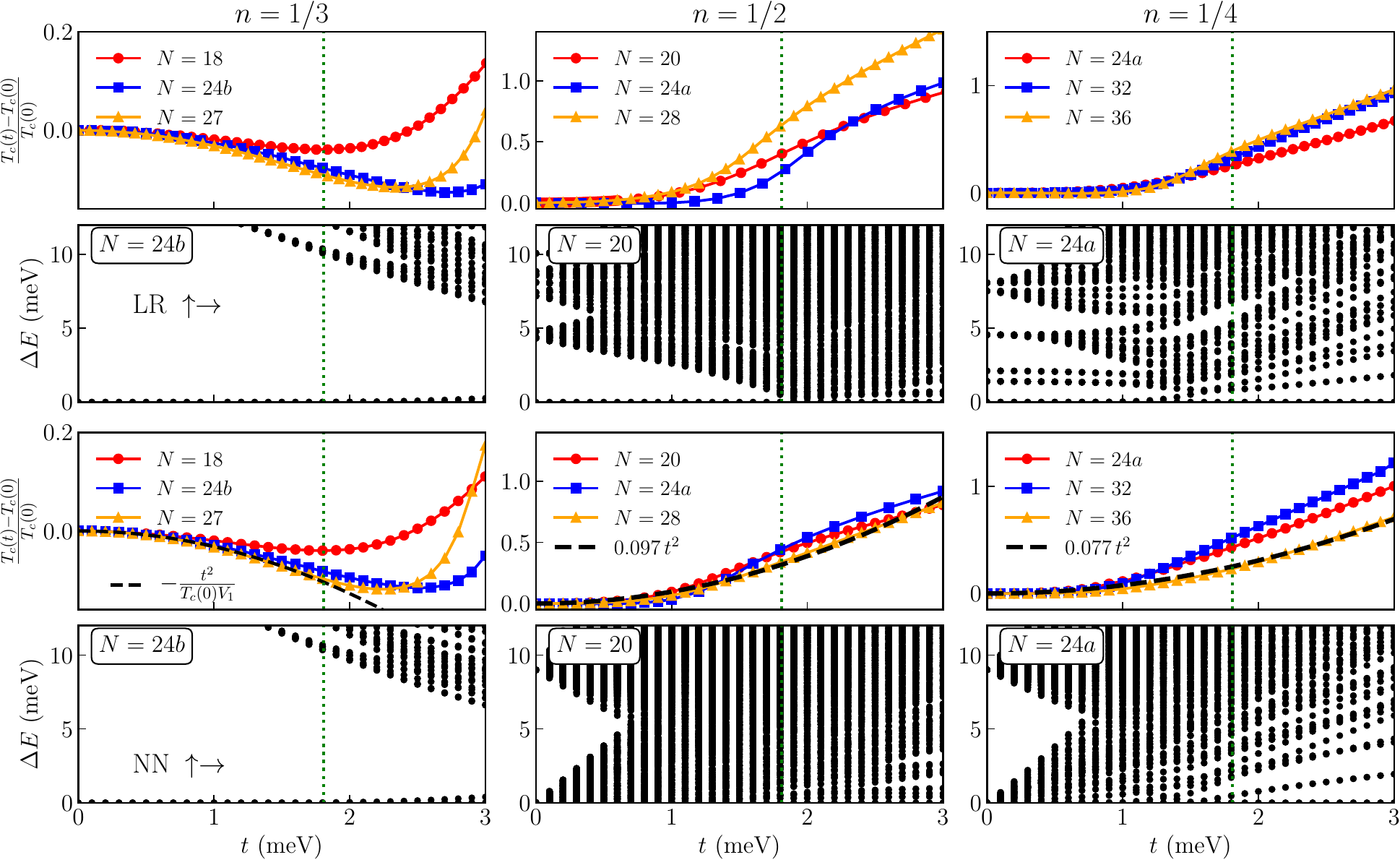}
\caption{Normalized critical temperature $(T_c (t) - T_c(0))/T_c(0)$ as a function of the hopping amplitude $t$ for three different fillings. From left to right, the panels correspond to $n = 1/3$, $n = 1/2$, and $n = 1/4$. The symbols show numerical results obtained from exact diagonalization or the finite-temperature Lanczos method for representative system sizes ($N$). Results are shown for the LR model and NN model (first and third panel from top respectively). 
The second and fourth panels from the top show the many-body spectra as a function of $t$ for the LR and NN models on a representative system size, after subtracting out the ground state energy for a given $t$. Dotted lines correspond to $t=1.8$ meV. For $n = 1/3$, the 
black dashed line represents the quadratic suppression predicted by the analytic expression $(T_c (t) - T_c(0))/T_c(0) = -t^2/(T_c(0) V_1)$, and for the other cases it represents a quadratic fit. }
\label{fig:tc_vs_t}
\end{figure*}

\textit{Finite temperature melting of GWC at various densities}
In Fig.~\ref{fig:specific_heat}, we present the specific heat $C_v /N$ (heat capacity per site), versus temperature $T$ for a representative system size for $n=1/3, 1/2$ and $1/4$, for 
different values of $t$ (taken to be NN) in the range of $0$ to $3$ meV. We have considered two sets of models: The full (screened) LR model with $\epsilon=4.67$ and $d/a=10$ taking $a=7.98$ nm (top panels) and the NN model with $V_1=9$ meV (bottom panels).  
[The effective mapping of the LR to NN models was carried out for $n=1/3$ by matching the specific heat or energetics, a simplified form of downfolding.~\cite{Schuler2013, Changlani_downfolding_2015, Zheng_downfolding_2018}]

The qualitative results for both models are similar, but there are quantitative differences especially for $n=1/2$ and $n=1/4$. 
The vertical dashed lines mark the temperatures $T_c$ where $C_v /N$ reaches its maximum. 
We have shown the 
location 
both for $t=0$ (classical case) and $t=1.8$ meV (which is supposed to closely represent the real material). 
In the GWC phase this maximum must be interpreted as the temperature where charge order is destroyed, and in the Fermi liquid (FL) phase (which is not conventionally ordered) it corresponds to a 
characteristic scale associated with the kinetic energy (above this scale the low temperature FL is disordered). 

Notwithstanding the broadness of the peak/bumps in the specific heat, due to the small system sizes, some trends are quite striking. For the case of $n=1/3$ this feature clearly moves to slightly lower values on increasing $t$ from zero in the GWC phase. [For large enough $t$, the ground state eventually transitions to the FL and for this parameter regime we find that this feature increases 
roughly linearly with $t$, for large enough $t$.]
This \textit{decrease} for small $t$ is consistent with the qualitative notion that quantum fluctuations aid thermal fluctuations, and so a smaller physical temperature suffices to melt the GWC to a charge disordered phase. The NN and LR models are largely in quantitative agreement as well, which confirms the picture developed in earlier work~\cite{kumar_wigner_2025}. 

However in the $n=1/2$ and $n=1/4$ cases, the movement of this feature is exactly \textit{the opposite} of that for the $n=1/3$ case: i.e. $T_c(t)$ continues to \textit{increase} with $t$, showing no sign of a decrease anywhere. This suggests that the GWC has become thermally stabler on the introduction of $t$ (assuming $t$ is small enough to keep the ground state as GWC). Thus the usual (and often invoked) qualitative argument that ``quantum fluctuations aid thermal ones" is by no means rigorous, and needs to be revisited.

We take a closer look at these trends by plotting in Fig.~\ref{fig:tc_vs_t} the normalized location of the specific heat maximum with $t$ i.e. $\frac{\delta T_c (t)}{T_c(0)} = \frac{T_c(t) - T_c(0)}{T_c(0)}$, while keeping the strength of the Coulomb part fixed in Eq.~\eqref{eq:model_ham}. [The normalization by $T_c(0)$ is invoked to compare different system sizes on the same scale]. For $n=1/3$ we show three system sizes $N=18, 24$ and $27$ which agree very well in the perturbative limit, showing a decrease that goes as $t^2$. However, there are differences at larger $t$, especially near the quantum phase transition between GWC to FL. The $27$ site hexagonal shaped cluster (which has the symmetries of the thermodynamic limit) shows a very prominent dip in $T_c$ with $t $. Overall, this profile gives the impression of a ``critical fan" associated with second order quantum phase transitions. However, the nature of the $T=0$ quantum phase transition has been argued to be first order (for the NN case).~\cite{Zhou_PRL_2024} For $n=1/3$ we note that the location of this dip is around $V_1/t \approx 9/2.5 = 3.6$, which is consistent with the location of the $T=0$ critical point separating the GWC from the FL [For large $U/t$ and not restricting to polarized spins, a similar value can be inferred from Ref.~\cite{Zhou_PRL_2024} (see their large $U/t$ results)]. In contrast, for $n=1/4$ and $n=1/2$ there is no perceptible decline of $T_c$ at small $t$, instead it increases with $t$ with a $t^2$ dependence at low $t$. [On closer inspection, we find that for small $t$ one can not rule out a flat feature]. Both LR and NN models show the same qualitative trends even though their low energy properties are different, as we discuss below. For completeness we also discuss our results for the $n=2/5$ case in Appendix \AppD.

In the second and fourth row panels of Fig.~\ref{fig:tc_vs_t} we also plot the many-body spectra (subtracting out the ground state energy for a given $t$) of a representative cluster (that could be addressed with full diagonalization), restricting the plot to an energy window of 12 meV. The
dotted line corresponds to $t = 1.8$ meV. For $n=1/3$, for both the LR and NN models, the degeneracy in the GWC ground state is clear -- both harbor a $\sqrt{3} \times \sqrt{3}$ three-fold degenerate GWC ground state.
For the case of $n=1/2$ and $n=1/4$ the NN model at $t=0$ has a high exact degeneracy (for example for $n=1/2$ the stripes are degenerate with ``bent stripes") and for the LR model this is lifted in favor of stripes for $n=1/2$ (which has a 6-fold exact degeneracy in the thermodynamic limit) and the $2 \times 2$ triangular crystal for $n=1/4$. 
The ground states of the LR and models naturally also differ for finite $t$, prominently, there are low lying (gapless) states for the NN model. Previous work on the NN model for $n=1/2$ suggests the existence of a ``pinball metal" ground state with charge order.~\cite{Hotta_2006, Tocchio_PRL2014, Fratini_pinball}~\footnote{A more detailed study of the $n=1/2$ NN case will be presented elsewhere.} 

We focus on the significant reorganization of the many-body spectra past some ($n$-dependent) critical value of $t$. The location of this feature is associated with a critical point. While this is difficult to precisely pin down when the low energy spectrum is gapless, and when the system is small, our results should provide a reasonable ballpark estimate of where the GWC transitions to a FL.

The choice of $t=1.8$ meV places the system in the GWC phase for $n=1/3$, 
for $n=1/2$ the system is close to the GWC-FL boundary, and for $n=1/4$ it appears to be in the FL regime. Several possibilities arise from these observations, we list them here:
\begin{itemize}
\item The value of $t=1.8$ meV from theory is possibly too large to explain the occurrence of the $n=1/4$ GWC.
\item Alternatively, the renormalization of the effective NN interaction is strongly density ($n$) dependent -- its value is potentially larger than 9 meV for $n=1/2, 1/4$. Note that $V_1$ was previously estimated using the information for the $n=1/3$ case
~\cite{kumar_wigner_2025}.
\item Experimentally, $n=1/4$ is not a GWC -- it has not been directly imaged with STM (yet), and all the evidence for it is only from optical measurements.
\end{itemize}

Addressing which one of these holds would require more inputs from experiment -- we leave this to future work. The plots do however establish that for all densities shown, the experimentally relevant GWC is in a non perturbative regime of $t$, yet it is adiabatically connected (where applicable) to the perturbative regime i.e. small $t$. This calls for a development of a finite temperature perturbative approach to understand the origin of the dependence of $T_c$ on $t$, and we do so next.

\begin{figure}
    \centering
    \includegraphics[width=\linewidth]{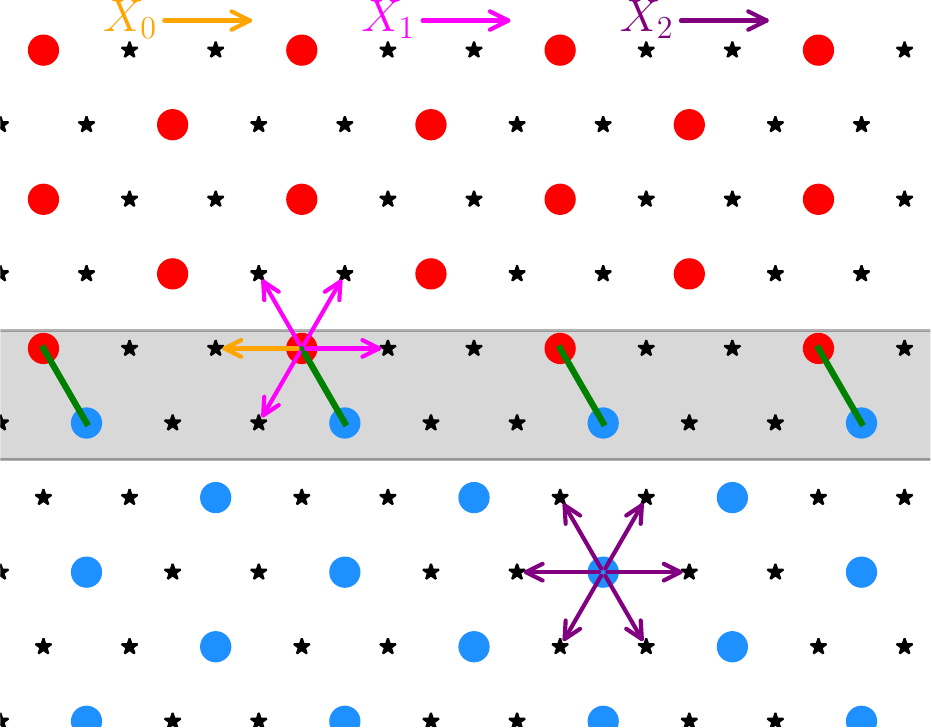}
    \caption{Schematic of a single domain wall (marked in grey) between two domains (colored ``red" and ``blue" corresponding to different sublattices) for $n=1/3$. Notice that a shear transformation (shifting the blue domain one lattice constant to the right) restores the ordered state, so that there is no density correction in the calculation of the domain wall energy. The arrows emanating from a particle indicate the various nearest-neighbor hops (based on the kinetic contribution). The colors of the arrows correspond to the $X_k$ contributions that enter the partition function, as explained in the text and appendix. We provide an analytic calculation for the change in domain wall energy when quantum effects are incorporated perturbatively in the NN classical model.
    }
    \label{fig:n_1b3_domains}
\end{figure}
\textit{The case of $n=1/3$:}
We consider $H = K + V $ where $K$ is the kinetic energy operator (hopping) and $V$ is the Coulomb piece, and treat the former as a perturbation in the small $t$ regime. 
We approximate the partition function $Z = \textrm{Tr} e^{-\beta H} = \textrm{Tr} e^{-\beta (V + K)}$ as (see Appendix \AppF),
\begin{equation}
Z \approx \sum_i e^{-\beta V_i} \Big( 1 + \sum_j X_{ij}(\beta)  \Big) \equiv \sum_i e^{-\beta V'_i}
\end{equation}
$|i \rangle$ is used to denote the many-body classical occupation number states i.e. eigenstates of $V$ and we have defined $V |i \rangle = V_i | i \rangle$ and $V'_i$ is its effective value once quantum effects have been incorporated.
The dimensionless quantity $X_{ij}(\beta)$ is defined as,
\begin{eqnarray}
X_{ij} (\beta) & \equiv & \frac{\beta |K_{ij}|^2}{\Delta_{ij}} \Big( 1 - \Big( \frac{1 - e^{-\beta \Delta_{ij}}}{\beta \Delta_{ij}}\Big) \Big)
\end{eqnarray}
where $\Delta_{ij}=V_j - V_i$. Assuming NN $t$ and under additional approximations, we get
\begin{equation}\label{V'}
V'_i=V_i-t^2\sum_j \frac{1}{\Delta_{ij}}\Bigg(1-\left(\frac{1-e^{-\beta\Delta_{ij}}}{\beta\Delta_{ij}}\right)\Bigg)
\end{equation}
$\Delta_{ij}$ that arise in the sum over $j$ can be positive, negative or zero-- the summand equals $\beta/2$ when $\Delta_{ij}=0$. 

Since the NN and (experimentally relevant) LR model have many similarities for $n=1/3$, we simplify our analytic calculation by working with the former. 
The $n=1/3$ ordered state has the 3-state Potts model symmetry, with three symmetry-related domains and equal (by symmetry) domain-wall energies. The critical ordering temperature is therefore proportional to the energy scale associated with the energy/length of a domain wall. We assume that quantum effects only change the energy/length of domain walls. Strictly speaking, we need to estimate this near $T_c$ but a rigorous calculation is tricky and instead we estimate its value at $T=0$. Despite this limitation, we find the resultant analytic expression to be remarkably consistent with our numerical FTLM results in the perturbative regime, as shown in Fig.~\ref{fig:tc_vs_t}.

As Fig.~\ref{fig:n_1b3_domains} shows, the lowest-energy ($T=0$) domain walls are aligned with the edges of the underlying triangular lattice, not the triangle edges of the $\sqrt{3}\times\sqrt{3}$ lattice of the GWC. For the purely classical case, the domain wall energy per length is $\mu=\frac{V_1}{3a}$, where $a$ is the triangular lattice constant, and $3a$ is the repeat-distance of the domain wall. Corrections to $\mu$ come from differences in the sum over $j$ in \eqref{V'} for the pairs of charges that have spacing $a$ along the domain wall, and we work out the details in Appendix \AppF. We find,

\begin{equation}
\delta\mu=\frac{t^2}{V_1}\, \frac{f(\beta V_1)}{3a},
\end{equation}
where
\begin{equation}\label{f(w)}
f(w)=-w -8\left(1-\frac{1-e^{-w}}{w}\right)+6\left(1-\frac{1-e^{-2w}}{2w}\right) .
\end{equation}
Apart from $\beta V_1$ in the range $[0,0.5]$, $f$ is negative and therefore \textit{lowers} the domain-wall energy and $T_c$. Most of that behavior comes from the $-w$ term, which in turn comes from the feature that charges in the domain wall can hop without any potential-energy penalty in the NN 
model.
Since the domain-wall energy sets the scale of $T_c$, we get,
\begin{eqnarray}
\frac{\delta T_c(t)}{T_c(0)}&=&\frac{\delta \mu}{\mu} =\frac{t^2}{V_1^2}f(\beta V_1).
\label{eq:second_order}
\end{eqnarray}
Using $V_1\approx 5 t$ and $w=V_1/T_c\approx 2.8$, the shift in $T_c$ comes out to about -11\%, which is the ballpark of what we see numerically.

\textit{The cases of $n=1/2, 1/4$:} As seen earlier for these cases, the NN model does not harbor a conventional insulating GWC ground state, instead there is strong evidence of gapless excitations, consistent with previously reported phase diagrams (for $n=1/2$~\cite{Hotta_2006, Tocchio_PRL2014, Fratini_pinball}). To drive the system towards a GWC, one would need (at the very least) a NNN interaction. Assuming that this modified minimal Hamiltonian realizes a GWC, the question now is how does it melt, and what impact does $t$ have on it. The argument we developed for $n=1/3$ crucially depends on the existence of a second order phase thermal transition, and hence is not applicable to these cases, which have been previously argued to be first order for $t=0$~\cite{mak_xu2020correlated}.

At $T_c$, irrespective of the order of the transition, the free energy of the ``ordered phase" (``o", low temp side) and the ``disordered phase" (``d", high temp side) must match exactly i.e. we must have 
$E_{o}(t) - T_{c}(t) S_{o}(t) = E_{d}(t) - T_{c}(t) S_{d}(t)$ where $E, S$ represent the thermodynamic internal energy and entropy respectively. Using $\delta$ to represent the change in a quantity due to quantum effects, and ignoring $\delta S \delta T$ terms we get,
\begin{equation}
 \frac{\delta T_c (t)}{T_c(0)} \approx \frac{ \frac{(\delta E_d(0) - \delta E_o(0))}{T_c(0)} - (\delta S_d - \delta S_o)}{S_{d}(0) - S_{o}(0)}
 \label{eq:first_order}
\end{equation}
The classical entropy of the disordered phase is larger than the ordered phase and so the denominator is positive. The numerator can be positive in various cases including situations where $\delta E_d < \delta E_o $ but $\delta S_d - \delta S_o$ is sufficiently negative. Said physically: a decrease in the ``entropy jump" at the transition between the ordered and disordered state, due to quantum effects, is equivalent to making the ordered state more stable than the disordered state at $T_c$. Entropic stabilization of the ordered state means one would have to go to higher temperatures to melt it. The entropic stabilization mechanism can come in the form of quantum fluctuating stripes for $n=1/2$, or shear modes (translational displacement of a row/column of electrons) in the case of the $2 \times 2$ crystal for $n=1/4$. These fluctuations have to be spatial in nature for consistency with our calculations in which spin was suppressed. 

While a detailed microscopic mechanism needs to be developed on a case by case basis by analyzing the defects associated with the disordered state, one must admit the possibility that these can energetically and/or entropically yield a net positive $t^2$ scaling of $\delta T_c$ in the perturbative limit. To test these ideas we have fit our FTLM data in Fig.~\ref{fig:tc_vs_t} to $\alpha t^2$ and obtained very good fits. Unlike the case of $n=1/3$, the resulting $\alpha$ is now more sensitive to system geometry and size, because the classical entropies and $T_c(0)$ that enter Eq.~\eqref{eq:first_order} are now system/geometry specific, in sharp contrast to Eq.~\eqref{eq:second_order}. 

\textit{Conclusions:}
In summary, we conclude that while quantum fluctuations naturally soften the ground-state order parameter for GWCs, their impact on the transition temperature $T_c$ is highly filling-dependent. For $n = 1/3$ the introduction of hopping slightly reduces $T_c$ with $t$, consistent with the intuitive expectation that quantum melting destabilizes crystalline order. In contrast, for fillings $n = 1/4$ and $1/2$,  quantum fluctuations \textit{increase} the melting temperature by as much as 30--40\%, at least on the system sizes studied here. This counterintuitive enhancement can be understood as arising from energetic and entropic contributions to melting.

We have developed a finite temperature perturbation theory (Appendix \AppE) to pursue an analytic understanding for why $T_c$ generally scales as $t^2$ in the perturbative limit (with either sign) -- this is supported by strong numerical evidence from exact diagonalization and finite temperature Lanczos. In the $n=1/3$ case, we carried out a detailed microscopic calculation on the NN model (Appendix \AppF) and related the \textit{reduction} in transition temperature with increasing quantum fluctuations to the \textit{reduction} in domain wall energy, as is expected for a second order thermal transition. This revealed the existence of multiple contributions to $\delta T_c (t) $ that scale as $t^2$, all sensitively dependent on how melting is modeled classically. For the $n=1/2$ and $n=1/4$ case, the NN classical model hosts a highly degenerate manifold of states, which is inadequate to model a GWC ground state. Additionally, the impact of quantum effects on the entropic contribution is needed to model a first order thermal transition, crucial for any mechanism where the transition temperature \textit{increases} with increasing quantum fluctuations. It would also be useful to evaluate the free energy within a ``order-by-disorder" type framework~\cite{Villain_1980, Henley_obd_1989} to develop further analytic understanding near the melting transition in these cases. 

These findings highlight that the thermal stability of GWCs cannot be inferred solely from the magnitude of the zero-temperature order parameter; instead, it reflects a subtle interplay between quantum fluctuations and configurational entropy. Importantly, the inclusion of quantum effects brings the theoretical $T_c$ closer to experimental values for several fillings, narrowing discrepancies that were as large as 50\% in the classical model. The role of quenched disorder~\cite{hammam2025} may also be important towards a conclusive quantitative explanation of the experiment, and we leave such an exploration to future work. We also note that in the context of Wigner crystals in the continuum, recent work has provided arguments for quantum effects being instrumental in increasing the melting temperature relative to what might be expected classically~\cite{Jain_2025}. 

The existence of classical-quantum shifts in either direction is not restricted to GWCs, and have previously been observed in other contexts  -- e.g. in quantum and classical simulations of magnetic pyrochlores~\cite{Changlani_YbTO, Scheie2017}, however, much remains to be understood in these systems~\cite{hallas2018experimental}.
While the microscopic mechanism of such differences needs to be reviewed on a case-by-case basis, our work clarifies some of the guiding principles that are rooted in thermodynamics. Importantly, our calculations of the material specific model extended Hubbard Hamiltonian enable predictions that can be tested in future experiments where the bandwidth can be tuned. It would also be illuminating to theoretically explore the role of quantum statistics on the melting temperature -- a bosonic version of the GWC state is the triangular supersolid.~\cite{Melko_PRL_2005, Boninsegni_PRL_2005,Heidarian_PRL_2005} 

\textit{Acknowledgments:}
We thank V. Dobrosavljevic, K. Yang, K.F. Mak, O. Tchernyshyov, S. Sachdev, S. Shastry, I. Esterlis, T. Scaffidi, C. Chamon, S. Zhang, and C. Lewandowski for useful discussions and for pointing us to the relevant literature and results. The simulations were performed
using a combination of the QuSpin \cite{quspin} library and our custom codes. We thank the Research Computing Center (RCC) and the Planck cluster at Florida State University for computational resources. This work also used Bridges-2 at Pittsburgh Supercomputing Center through allocation PHY240324 (Towards predictive modeling of strongly correlated quantum matter) from the Advanced Cyberinfrastructure Coordination Ecosystem: Services and Support (ACCESS) program, which is supported by U.S. National Science Foundation grants $\#$2138259, $\#$2138286, $\#$2138307, $\#$2137603, and $\#$2138296. We acknowledge support from the National High Magnetic Field Laboratory (NHMFL). The NHMFL is supported by the National Science Foundation through NSF/DMR-2128556 and the state of Florida. S.S. and H.J.C acknowledge funding from National Science Foundation Grant No. DMR 2046570. A.K. was supported through a Dirac postdoctoral fellowship at NHMFL. 

\appendix
\renewcommand\thesection{\Alph{section}}
\section{APPENDIX A: Extracting the dielectric constant $\epsilon$ from classical Monte Carlo and experiments}
\label{app:A}

\begin{figure}
    \centering
    \includegraphics[width=1\linewidth]{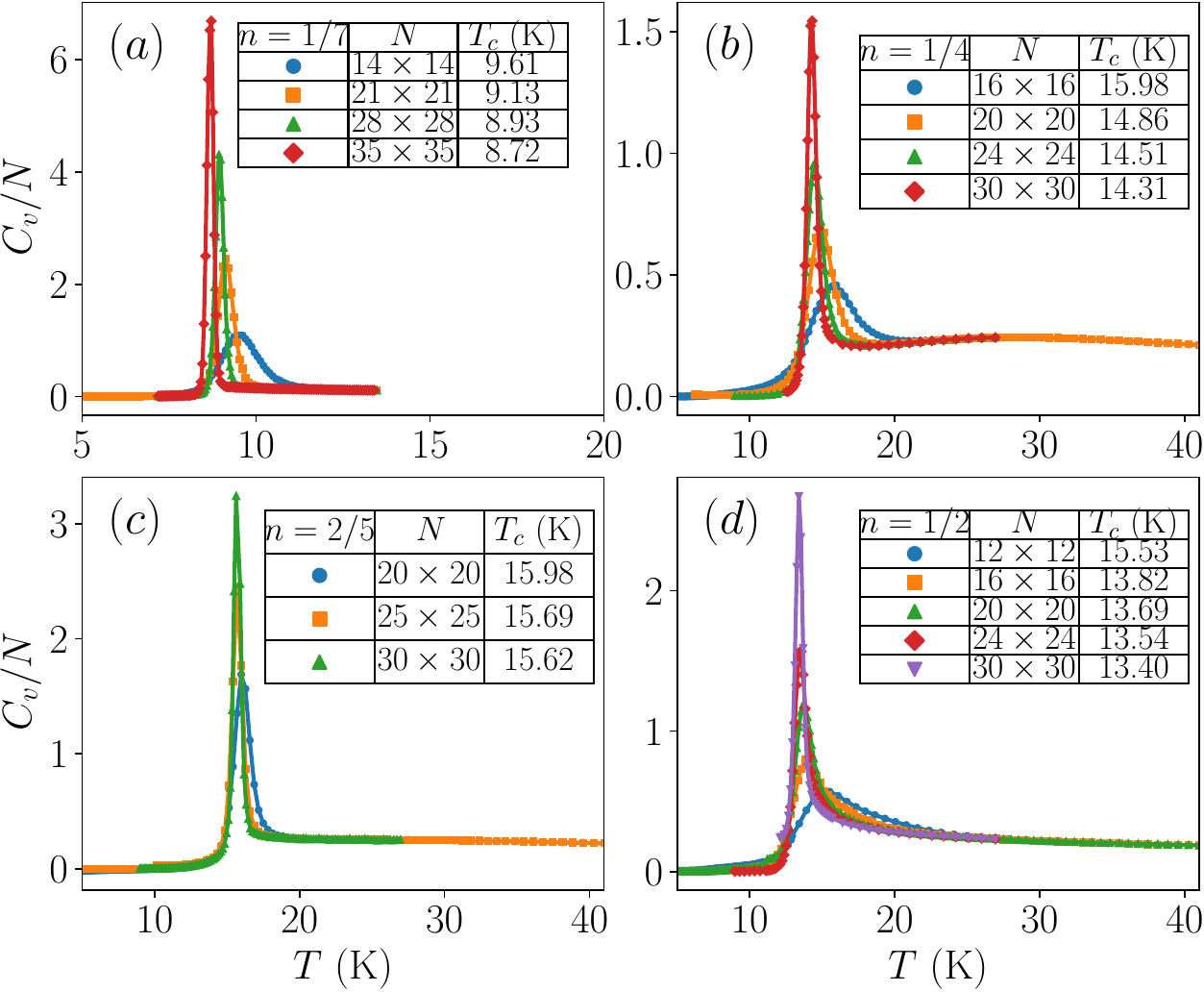}
\caption{
Specific heat (heat capacity per site in arbitrary units), $C_v/N$, as a function of temperature (in Kelvin) for various fillings and system sizes, obtained from classical Monte Carlo simulations using the replica method. Results are for the long range potential with $a=7.98$ nm and $d/a=10$. $\epsilon=4.67$ was obtained by fitting data obtained from the $n=1/3$ case (not shown here). Distinct markers denote different lattice sizes, as indicated in the tables within each panel. The transition temperature $T_c$ is identified from the peak position of $C_v$ and listed for each system size. All simulations were performed with $10^{8}$ Monte Carlo sweeps. (One sweep is defined a sequence of $N$ single particle moves, including the possibility of a replica exchange)}. 
\label{fig:classical_heat_capacity}
\end{figure}

As mentioned in the main text, we have refit the classical Monte Carlo data to the experimentally measured $T_c$, ignoring the reported error bars. The classical Monte Carlo was performed with the replica/parallel tempering method~\cite{Marinari_1992, Hukushima_1996}. We used $T_{min}=0.01$ to $T_{max}=0.1$ in dimensionless units, and 128 replicas, the temperatures were chosen according to a geometric series. We find this to be necessary to eliminate ``glassiness" at low temperature.

In Fig.~\ref{fig:classical_heat_capacity} we have shown our results for the specific heat for large system sizes to locate $T_c$ for each density. For each density, we take the $T_c$ (location of the specific heat maximum) for the largest size we simulated -- our estimate  accurately reflects the thermodynamic limit to an error of substantially less than 1 K. This is sufficiently small, for the scales of interest here -- the discrepancy we are attempting to explain in our study is 10-15 K i.e. at least an order of magnitude or two bigger than the error in our estimate.  

Like in previous work~\cite{mak_xu2020correlated}, we set the single parameter $\epsilon$ by rescaling our classical Monte Carlo data to exactly match the $T_c$ of 37 K reported in the experiment (hence the ``NA" for the error in Table~\ref{tab:Tc_eps4p69}). This procedure ignores the renormalization of the classical $T_c$ by quantum effects -- in previous work it was estimated that this would change the quantum $T_c$ by a few K and hence $\epsilon$ by a few percent. 

Given all these approximations, we find $\epsilon$ to be approximately $4.67$, instead of the previously reported value of $3.9$. This difference is small and this choice should not change the qualitative conclusions of previous work, nor the main results of the present study. We work with this revised number throughout.

\section{APPENDIX B:  Details of the finite temperature Lanczos Method (FTLM)}\label{app:B}

To compute the ground state and finite temperature properties, we employed both exact diagonalization and the finite temperature Lanczos method (FTLM) for the quantum models. In the FTLM approach, the partition function is expressed through stochastic sampling over a set of random vectors, combined with the tridiagonalization of the Hamiltonian in the basis of Krylov vectors. For each random vector, a Lanczos iteration is performed, yielding approximate eigenvalues and corresponding weights, which are then used to construct thermally averaged quantities.

We exploit lattice translational symmetries to access larger system sizes by block-diagonalizing the Hilbert space into distinct symmetry sectors. The FTLM procedure is carried out independently within each sector, and the contributions from all sectors are then combined with appropriate weights. The resulting expression for the thermal average of the energy is given in Eq.\eqref{eq:ftlm}. 
\begin{equation}\label{eq:ftlm}
\langle H \rangle_{\beta} =
\frac{
\displaystyle \sum_{\Gamma} \frac{N_\Gamma}{R}
\sum_{r=1}^{R} \sum_{j=1}^{M}
e^{-\beta E_j^{(r,\Gamma)}}
\left| \langle r \, | \, \psi_j^{(\Gamma)} \rangle \right|^2
\, E_j^{(r,\Gamma)}
}{
\displaystyle \sum_{\Gamma} \frac{N_\Gamma}{R}
\sum_{r=1}^{R} \sum_{j=1}^{M}
e^{-\beta E_j^{(r,\Gamma)}}
\left| \langle r \, | \, \psi_j^{(\Gamma)} \rangle \right|^2
}
\end{equation}

Here, in Eq.~\eqref{eq:ftlm}, $\beta=1/T$ is the inverse temperature (using natural units $k_B=1$), $N_{\Gamma}$ is the dimension of the Hilbert space in the symmetry sector $\Gamma$, and $R$ is the number of random initial vectors used for stochastic sampling. The state $|r\rangle$ represents a random initial vector within the symmetry sector. The quantities $E_j^{(r,\Gamma)}$ are the eigenvalues of the Lanczos tridiagonal matrix generated from $|r\rangle$ in the sector $\Gamma$, and $|\psi_j^{(r,\Gamma)}\rangle$ are the corresponding Lanczos eigenvectors.

In each symmetry sector, we used up to $R = 200$ random initial vectors for stochastic averaging and a maximum Krylov subspace dimension of $M = 200$. We have verified that our results have well converged over the temperature range of interest.

We also calculate $\langle H^2 \rangle_{\beta}$ and compute the heat capacity as,
\begin{equation}
C_V = \frac{\langle H^2 \rangle_{\beta}  - {\langle H \rangle^2_{\beta}} } {T^2}
\end{equation}

\begin{figure}
    \centering
    \includegraphics[width=1\linewidth]{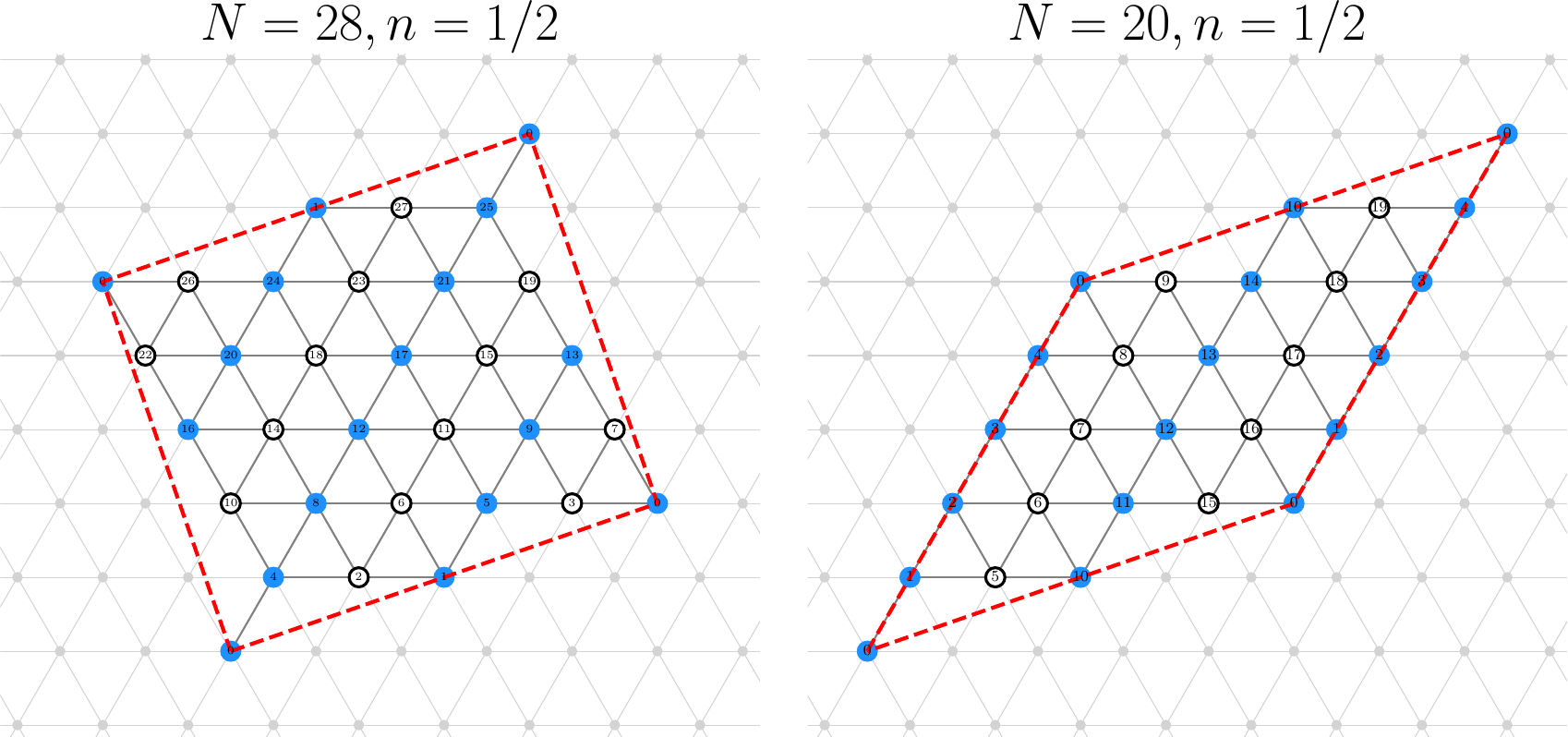}
    \includegraphics[width=1\linewidth]{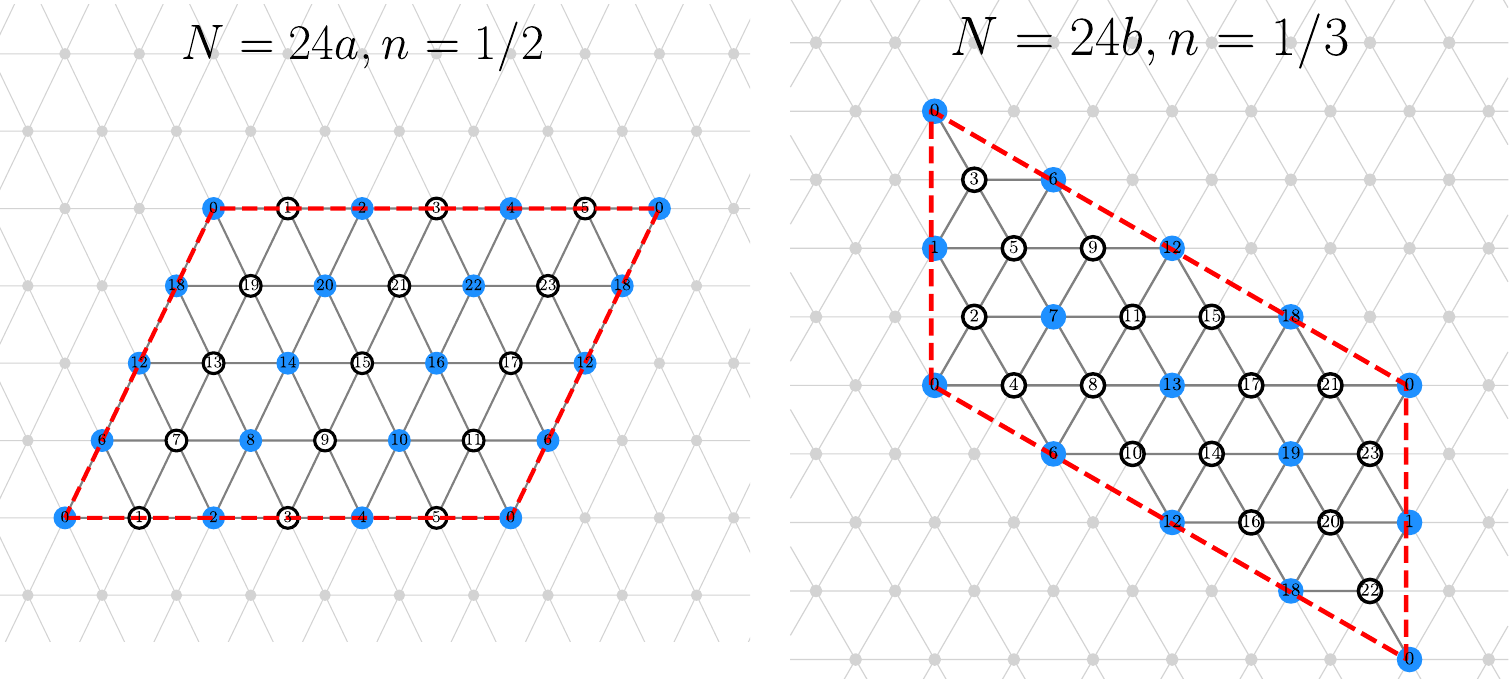}
    \caption{The top panel shows two geometries (28-site and 20-site) that support stripe configurations for $n=1/2$, indicated by filled blue circles. The bottom panel shows the two 24-site geometries, which accommodates the stripes ($n=1/2$) and $\sqrt{3} \times \sqrt{3}$ crystal structure ($n=1/3$).}
    \label{fig:lattice_geometries}
\end{figure}

\section{APPENDIX C: Cluster geometries used for exact diagonalization and FTLM simulations.
}\label{app:C}
We performed ED and FTLM simulations on several finite-size lattice geometries with periodic boundary conditions. In addition to parallelogram-type clusters such as 
24 sites ($6 \times 4$, which we call $24$a), 32 sites ($8 \times 4$), and 36 sites ($6 \times 6$), we employ other special geometries ($N=20$, $24b$, $28$) shown in Fig.~\ref{fig:lattice_geometries}. The use of multiple cluster shapes and sizes allows us to estimate finite-size effects and test the robustness of the results. We find that the qualitative trends in the transition temperature are consistent across the systems studied in this work.

\section{APPENDIX D: Results for the $n=2/5$ GWC}
\label{app:D}

\begin{figure}[]
    \centering
 
    \includegraphics[width=\linewidth]{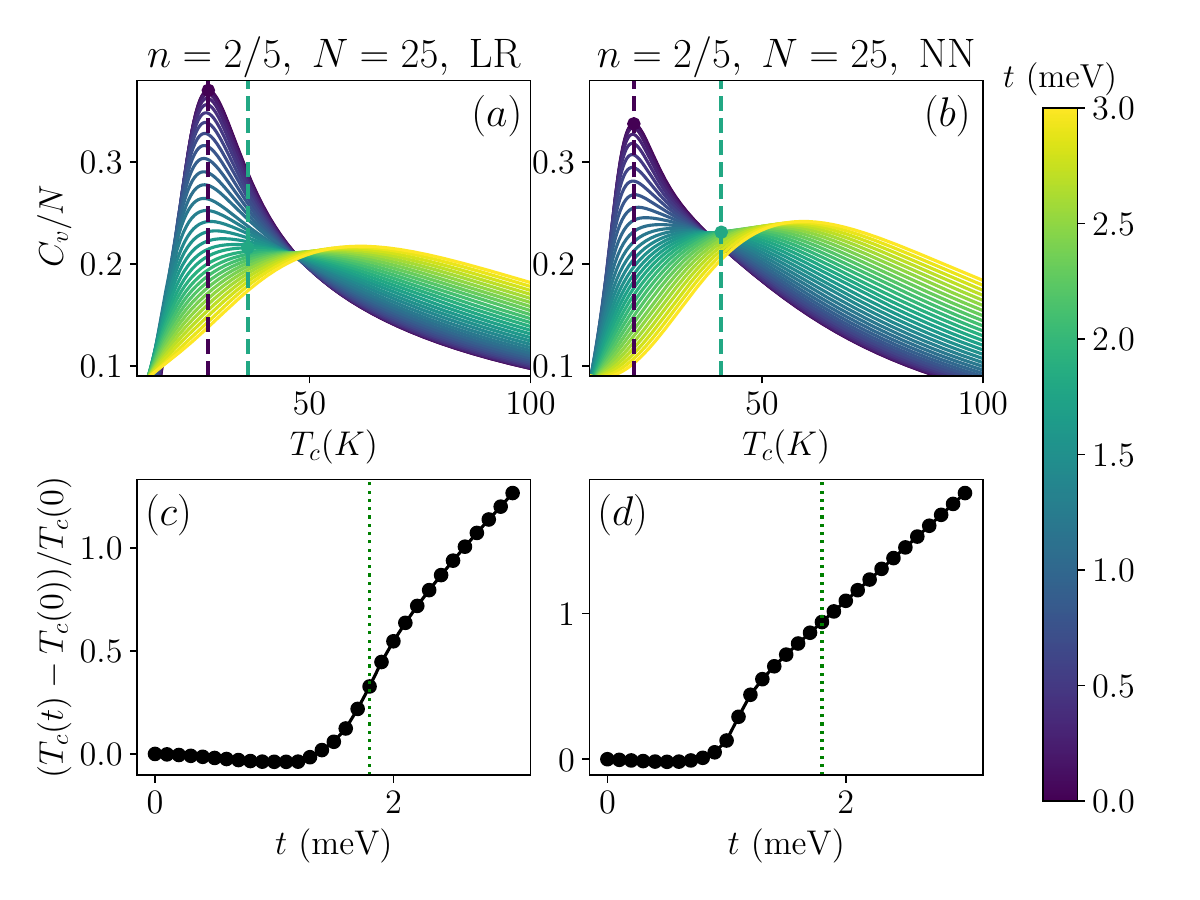}
               \includegraphics[width=0.6\linewidth]{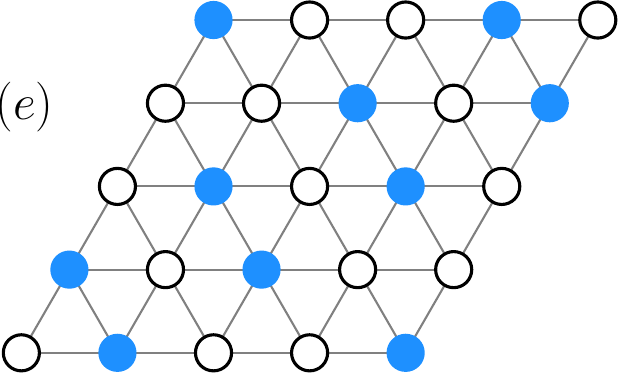}

    \caption{Temperature dependence of the specific heat (heat capacity per site in arbitrary units), $C_v/N$, for filling fraction $n = 2/5$ (top panels, (a,b)), and the normalized critical temperature $(T_c(t)-T_c(0))/T_c(0)$ as a function of hopping amplitude $t$ (middle panels, (c,d)). The simulations are performed using FTLM for the indicated system size $N = 5 \times 5$. Results are shown for both the long-range (LR) and nearest-neighbor (NN) models. The bottom panel (e) displays the classical configuration corresponding to $n=2/5$ for the $5 \times 5$ lattice considered.}
    \label{fig:2b5_crystal}
\end{figure}

In the main text we showed quantum mechanical calculations for $n=1/3,1/2$ and $1/4$. In Fig.~\ref{fig:2b5_crystal} we show results for the $n=2/5$ case for one representative size -- the $5\times5$ torus, with 10 particles (in the spin polarized sector). The specific heat data shows a small, but observable, reduction in $T_c$ with $t$ in the perturbative regime of the GWC phase, i.e. near $t=0$. While the LR and NN differ in quantitative details, they show similar qualitative trends. The figure also shows a schematic for the GWC for $n=2/5$.

\section{APPENDIX E: Effective classical Hamiltonian from finite temperature perturbation theory}\label{app:E}

To develop the finite temperature perturbation theory for the partition function, consider a general series expansion for the operator $e^{X+Y}$ in the limit that $Y$ is much smaller than $X$ (note $X$ and $Y$ are operators). Using the exact formula,
\begin{equation}
e^{X+Y} = \textrm{lim}_{N \rightarrow \infty} \Big( 1 + \frac{X+Y}{N} \Big)^N
\end{equation}
gives the leading order term to be $e^X$. Identifying the linear in $Y$ terms, and with a little bit of algebra (involving the conversion of discrete sum to integral), the first order term can be shown to be
\begin{eqnarray}
 \int_{s_1=0}^{s_1=1} e^{s_1 X} Y e^{(1-s_1)X} ds_1
\end{eqnarray}
A similar procedure for the second order term yields,
\begin{eqnarray}
 \int_{s_1=0}^{s_1=1} \int_{s_2 = 0}^{s_2 = 1 - s_1} e^{s_1 X} Y e^{s_2 X} Y e^{(1-s_1-s_2)X}  ds_2 ds_1
\end{eqnarray}
For the Hamiltonian $H=V+K$ identify $X=V$ and $Y=K$, the potential and kinetic parts respectively. The objective is to evaluate the partition function
$
Z = \textrm{Tr} e^{-\beta H} = \textrm{Tr} e^{-\beta (V + K)}
$
approximately using the formulae developed above. 

Using $|i \rangle$ to denote the many-body classical occupation number states i.e. eigenstates of $V$, and defining $V |i \rangle = V_i | i \rangle$ we get
\begin{equation}
Z_0 = \textrm{Tr} (e^{-\beta V}) = \sum_i e^{-\beta V_i} 
\end{equation}
Now since $K$ is completely off diagonal we get no contributions to the trace from it i.e.,
\begin{equation}
Z_1 = \sum_i \int_{s_1=0}^{s_1=1} e^{-\beta V_i} (-\beta) \langle i | K | i \rangle = 0  
\end{equation}
The second order term is,
\begin{widetext}
\begin{eqnarray}
Z_2 = \sum_{i,j} \int_{s_1=0}^{s_1=1} \int_{s_2 = 0}^{s_2 = 1 - s_1} e^{-s_1 \beta V_i} (-\beta)^2 |K_{ij}|^2  e^{-\beta s_2 V_j} e^{-\beta (1-s_1-s_2) V_i} ds_1 ds_2 
\end{eqnarray}
\end{widetext}
where we have introduced the identity $\sum_j |j \rangle \langle j| = 1$ and defined $K_{ij} \equiv \langle i | K | j \rangle$. Defining $\Delta_{ij} \equiv V_j - V_i$ we get,
\begin{widetext}
\begin{eqnarray}
Z_2 &=& \sum_{i,j} \int_{s_1=0}^{s_1=1} \int_{s_2 = 0}^{s_2 = 1 - s_1} e^{-\beta V_i} (-\beta)^2 |K_{ij}|^2 e^{-\beta s_2 \Delta_{ij}} ds_2 ds_1 \\
    &=& \sum_i e^{-\beta V_i} \sum_j \frac{\beta |K_{ij}|^2}{\Delta_{ij}} \Big( 1 - \Big( \frac{1 - e^{-\beta \Delta_{ij}}}{\beta \Delta_{ij}}\Big) \Big)
\end{eqnarray}
\end{widetext}

Assembling all the results we get,
\begin{equation}
Z \approx Z_0+Z_1+Z_2 = \sum_i e^{-\beta V_i} \Big( 1 + \sum_j X_{ij}(\beta)  \Big)
\end{equation}
where the dimensionless quantity $X_{ij}(\beta)$ is defined as,
\begin{eqnarray}
X_{ij} (\beta) & \equiv & \frac{\beta |K_{ij}|^2}{\Delta_{ij}} \Big( 1 - \Big( \frac{1 - e^{-\beta \Delta_{ij}}}{\beta \Delta_{ij}}\Big) \Big)
\end{eqnarray}
We then define the effective potential as, 
\begin{equation}
 \sum_i e^{-\beta V^{'}_i} \equiv \sum_i e^{-\beta V_i} \Big( 1 + \sum_j X_{ij}(\beta)  \Big) 
\end{equation}
Thus,
\begin{equation}
V^{'}_i = V_i - \frac{1}{\beta} \ln \Big( 1 + \sum_j X_{ij}(\beta) \Big)
\end{equation}
Assuming $\sum_j X_{ij}(\beta)$ is small, we get,
\begin{equation}
V^{'}_i \approx V_i - \frac{\sum_j X_{ij}(\beta)}{\beta}
\end{equation}
\section{APPENDIX F: Application of the effective classical Hamiltonian to the $n=1/3$ case}\label{app:F}

The expansion of the effective Hamiltonian in powers of $t$ (hopping) is a regular power series in that parameter. In particular, for the NN model we have,
\begin{equation}\label{eq:V'}
V'_i=V_i-t^2\sum_j \frac{1}{\Delta_{ij}}\Bigg(1-\left(\frac{1-e^{-\beta\Delta_{ij}}}{\beta\Delta_{ij}}\right)\Bigg)
\end{equation}
is the exact correction up to $O(t^2)$ irrespective of the values of $\Delta_{ij}$ ---positive, negative or zero---that arise in the sum over $j$ (the summand equals $\beta/2$ when $\Delta_{ij}=0$). 
As stated in the main text, the $n=1/3$ ordered state has the 3-state Potts model symmetry, with three symmetry-related domains and equal (by symmetry) domain-wall energies. The critical ordering temperature is therefore proportional to a single energy scale---the energy/length of a domain wall. We evaluate the energy of a domain wall using snapshots of $T=0$ states, but then, as we explain below, invoke the temperature-dependent effective Hamiltonian which includes quantum effects. We find the energy of the domain wall for finite $t$ to be lower than the classical case ($t=0$), for the physically relevant parameter regime, thereby explaining why the classical $T_c$ must be higher than the quantum one.
The lowest-energy ($T=0$) domain walls are aligned with the edges of the underlying triangular lattice, not the triangle edges of the $\sqrt{3}\times\sqrt{3}$ lattice of the Wigner crystal. The main text shows a sketch of a domain wall (see Fig.~\ref{fig:n_1b3_domains}); the energy of a domain wall can be thought of as the energy difference between two states: one with a single domain wall ($E_D$) that splits the entire system into two halves, i.e. half the GWC is on the $A$ sublattice and half of it is on the $B$ sublattice, and the other that has no domain wall ($E_A$), i.e. the GWC is purely on the $A$ sublattice. Thus in the classical case for the NN model it is easy to to see,
\begin{equation}
 \mu = \frac{E_D - E_A}{L_{wall}} = \frac{V_1}{3a}  
\end{equation}
where $L_{wall}$ is the length of the domain wall, $V_1$ is the near-neighbor interaction, $a$ is the lattice constant, and $3a$ is the repeat-distance of the domain wall. 

Now let us reconsider the above calculation for non zero $t$ by evaluating energies $E'_D$ and $E'_A$. It will be convenient to define
\begin{equation}
X_k=\frac{1}{k}\Bigg(1-\left(\frac{1-e^{-k w}}{k w}\right)\Bigg),
\label{eq:x_eqn}
\end{equation}
where $w=\beta V_1$ and $k=0,1,2$ covers all the cases that arise in the sum over states $j$ (in the situation we consider here). Note we can take the limit of $k \rightarrow 0$ to get the result $X_0=w/2$. 

\begin{figure}[]
    \centering
    \includegraphics[width=\linewidth]{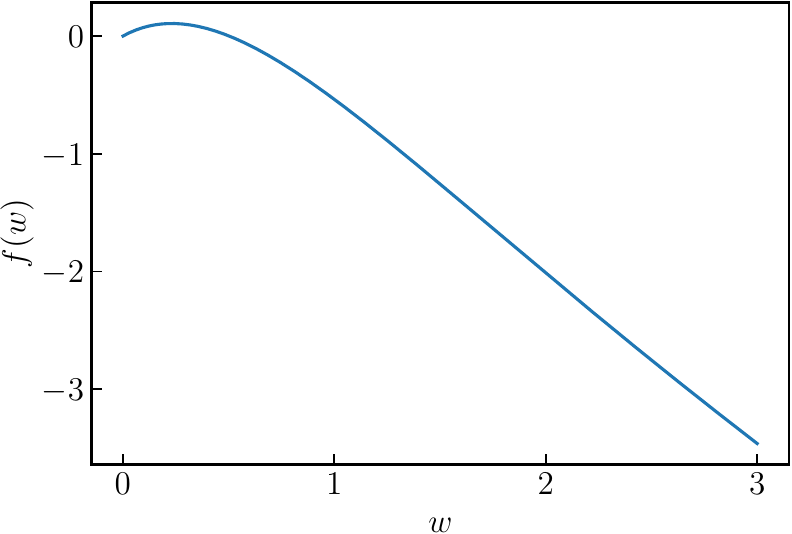}
\caption{A plot of the function $f(w)$, as defined in Eq.~\eqref{f(w)}, giving the temperature dependence of the quantum correction to the domain-wall energy.}
\label{fig:f_fn}
\end{figure}

To calculate $E'_D$ consider a ``red" particle (i.e. in sublattice $A$) in the domain wall, that can hop to any one of its 5 neighbors. For 4 of these hops the energy of the new configuration increases by an additional $2V_1 - V_1 = V_1$ (i.e. two nearest neighbor costs are paid, but one is lost) which corresponds to the case of $k=1$ in Eq.~\eqref{eq:x_eqn}. For one of the 5 hops the energy of the new and old configurations is exactly the same, which corresponds to the case of $k=0$. Repeating the same calculation for the ``blue" particles  (i.e. in sublattice $B$) in the domain wall yields exactly the same contributions. 
The contributions from hops of the other particles not in the domain wall are referred to as ``rest" and do not need to be evaluated since they will eventually drop out in the calculation of $\mu'$. Thus we have,
\begin{equation}
E'_D = N_w V_1 -\frac{2 N_w t^2}{V_1} \Big(4 X_1 + X_0 \Big) + \textrm{rest}
\label{eq:edprime}
\end{equation}
where $N_w$ is the number of particles of one sublattice that are part of the domain wall. Now consider the configuration with no domain wall. Each of the particles can move to any of the 6 nearest neighbors -- the energy difference of the newly created configuration with respect to the original one is $2V_1 -0 = 2V_1$. Doing this for $2 N_w$ particles we get,  
\begin{equation}
E'_A = 0 -\frac{2 N_w t^2}{V_1} 6 X_2 + \textrm{rest}
\label{eq:eaprime}
\end{equation}
where the ``rest" is identical to that in Eq.~\eqref{eq:edprime}
Assembling all our results together gives,
\begin{equation}
\mu' = \frac{E'_D - E'_A}{L_{wall}} = \frac{V_1}{3a} - 2\frac{t^2}{V_1} \frac{\Big(X_0 + 4 X_1 - 6 X_2 \Big)}{3a}   
\end{equation}
where we have used $N_w/L_{wall} = 1/3a$. 
Evaluating the ``quantum correction" to the domain wall energy per unit length yields,
\begin{equation}
\delta\mu= \mu' - \mu = \frac{t^2}{V_1}\, \frac{f(\beta V_1)}{3a},
\end{equation}
where
\begin{equation}\label{f(w)}
f(w)=-w -8\Bigg(1-\frac{1-e^{-w}}{w}\Bigg)+6\Bigg(1-\frac{1-e^{-2w}}{2w}\Bigg)
\end{equation}
This function is plotted in Fig.~\ref{fig:f_fn}. Apart from $\beta V_1$ in the range $[0,0.5]$, $f$ is negative and therefore \textit{lowers} the domain-wall energy and $T_c$. Most of that behavior comes from the term $-w$ in Eq.~\eqref{f(w)}, which in turn comes from the feature that charges in the domain wall can hop without any potential-energy penalty in the $V_1$ model (the case $k=0$).

We now provide a quantitative estimate of the reduction in $T_c$ in reference to experimentally relevant parameters. 
Since the domain-wall energy (per unit length) sets the scale of $T_c$,
\begin{eqnarray}
\frac{\delta T_c}{T_c(0)}&=&\frac{\delta \mu}{\mu} = \frac{t^2}{V_1^2}f(\beta V_1).
\end{eqnarray}
Using $V_1\approx 5 t$ and $w=V_1/T_c\approx 2.8$, (recall $V_1 \approx 9$ meV and $T_c \approx 37$ K), the shift in $T_c$ comes out to about -11\%. On an absolute scale, this is a shift of about 4 K, which is in the ballpark of what was seen in previous work.~\cite{kumar_wigner_2025}


\begin{thebibliography}{55}%
\makeatletter
\providecommand \@ifxundefined [1]{%
 \@ifx{#1\undefined}
}%
\providecommand \@ifnum [1]{%
 \ifnum #1\expandafter \@firstoftwo
 \else \expandafter \@secondoftwo
 \fi
}%
\providecommand \@ifx [1]{%
 \ifx #1\expandafter \@firstoftwo
 \else \expandafter \@secondoftwo
 \fi
}%
\providecommand \natexlab [1]{#1}%
\providecommand \enquote  [1]{``#1''}%
\providecommand \bibnamefont  [1]{#1}%
\providecommand \bibfnamefont [1]{#1}%
\providecommand \citenamefont [1]{#1}%
\providecommand \href@noop [0]{\@secondoftwo}%
\providecommand \href [0]{\begingroup \@sanitize@url \@href}%
\providecommand \@href[1]{\@@startlink{#1}\@@href}%
\providecommand \@@href[1]{\endgroup#1\@@endlink}%
\providecommand \@sanitize@url [0]{\catcode `\\12\catcode `\$12\catcode `\&12\catcode `\#12\catcode `\^12\catcode `\_12\catcode `\%12\relax}%
\providecommand \@@startlink[1]{}%
\providecommand \@@endlink[0]{}%
\providecommand \url  [0]{\begingroup\@sanitize@url \@url }%
\providecommand \@url [1]{\endgroup\@href {#1}{\urlprefix }}%
\providecommand \urlprefix  [0]{URL }%
\providecommand \Eprint [0]{\href }%
\providecommand \doibase [0]{https://doi.org/}%
\providecommand \selectlanguage [0]{\@gobble}%
\providecommand \bibinfo  [0]{\@secondoftwo}%
\providecommand \bibfield  [0]{\@secondoftwo}%
\providecommand \translation [1]{[#1]}%
\providecommand \BibitemOpen [0]{}%
\providecommand \bibitemStop [0]{}%
\providecommand \bibitemNoStop [0]{.\EOS\space}%
\providecommand \EOS [0]{\spacefactor3000\relax}%
\providecommand \BibitemShut  [1]{\csname bibitem#1\endcsname}%
\let\auto@bib@innerbib\@empty
\bibitem [{\citenamefont {Wigner}(1934)}]{Wigner_1934}%
  \BibitemOpen
  \bibfield  {author} {\bibinfo {author} {\bibfnamefont {E.}~\bibnamefont {Wigner}},\ }\bibfield  {title} {\bibinfo {title} {On the interaction of electrons in metals},\ }\href {https://doi.org/10.1103/PhysRev.46.1002} {\bibfield  {journal} {\bibinfo  {journal} {Phys. Rev.}\ }\textbf {\bibinfo {volume} {46}},\ \bibinfo {pages} {1002} (\bibinfo {year} {1934})}\BibitemShut {NoStop}%
\bibitem [{\citenamefont {Ceperley}\ and\ \citenamefont {Alder}(1980)}]{Ceperley_Alder_1980}%
  \BibitemOpen
  \bibfield  {author} {\bibinfo {author} {\bibfnamefont {D.~M.}\ \bibnamefont {Ceperley}}\ and\ \bibinfo {author} {\bibfnamefont {B.~J.}\ \bibnamefont {Alder}},\ }\bibfield  {title} {\bibinfo {title} {Ground state of the electron gas by a stochastic method},\ }\href {https://doi.org/10.1103/PhysRevLett.45.566} {\bibfield  {journal} {\bibinfo  {journal} {Phys. Rev. Lett.}\ }\textbf {\bibinfo {volume} {45}},\ \bibinfo {pages} {566} (\bibinfo {year} {1980})}\BibitemShut {NoStop}%
\bibitem [{\citenamefont {Nagara}\ \emph {et~al.}(1987)\citenamefont {Nagara}, \citenamefont {Nagata},\ and\ \citenamefont {Nakamura}}]{Nagara_1987}%
  \BibitemOpen
  \bibfield  {author} {\bibinfo {author} {\bibfnamefont {H.}~\bibnamefont {Nagara}}, \bibinfo {author} {\bibfnamefont {Y.}~\bibnamefont {Nagata}},\ and\ \bibinfo {author} {\bibfnamefont {T.}~\bibnamefont {Nakamura}},\ }\bibfield  {title} {\bibinfo {title} {Melting of the wigner crystal at finite temperature},\ }\href {https://doi.org/10.1103/PhysRevA.36.1859} {\bibfield  {journal} {\bibinfo  {journal} {Phys. Rev. A}\ }\textbf {\bibinfo {volume} {36}},\ \bibinfo {pages} {1859} (\bibinfo {year} {1987})}\BibitemShut {NoStop}%
\bibitem [{\citenamefont {Tanatar}\ and\ \citenamefont {Ceperley}(1989)}]{Tanatar_Ceperley_1989}%
  \BibitemOpen
  \bibfield  {author} {\bibinfo {author} {\bibfnamefont {B.}~\bibnamefont {Tanatar}}\ and\ \bibinfo {author} {\bibfnamefont {D.~M.}\ \bibnamefont {Ceperley}},\ }\bibfield  {title} {\bibinfo {title} {Ground state of the two-dimensional electron gas},\ }\href {https://doi.org/10.1103/PhysRevB.39.5005} {\bibfield  {journal} {\bibinfo  {journal} {Phys. Rev. B}\ }\textbf {\bibinfo {volume} {39}},\ \bibinfo {pages} {5005} (\bibinfo {year} {1989})}\BibitemShut {NoStop}%
\bibitem [{\citenamefont {Mehta}\ \emph {et~al.}(2013)\citenamefont {Mehta}, \citenamefont {Umrigar}, \citenamefont {Meyer},\ and\ \citenamefont {Baranger}}]{Mehta_2013}%
  \BibitemOpen
  \bibfield  {author} {\bibinfo {author} {\bibfnamefont {A.~C.}\ \bibnamefont {Mehta}}, \bibinfo {author} {\bibfnamefont {C.~J.}\ \bibnamefont {Umrigar}}, \bibinfo {author} {\bibfnamefont {J.~S.}\ \bibnamefont {Meyer}},\ and\ \bibinfo {author} {\bibfnamefont {H.~U.}\ \bibnamefont {Baranger}},\ }\bibfield  {title} {\bibinfo {title} {Zigzag phase transition in quantum wires},\ }\href {https://doi.org/10.1103/PhysRevLett.110.246802} {\bibfield  {journal} {\bibinfo  {journal} {Phys. Rev. Lett.}\ }\textbf {\bibinfo {volume} {110}},\ \bibinfo {pages} {246802} (\bibinfo {year} {2013})}\BibitemShut {NoStop}%
\bibitem [{\citenamefont {Smith}\ \emph {et~al.}(2024)\citenamefont {Smith}, \citenamefont {Chen}, \citenamefont {Levy}, \citenamefont {Yang}, \citenamefont {Morales},\ and\ \citenamefont {Zhang}}]{Zhang_2024}%
  \BibitemOpen
  \bibfield  {author} {\bibinfo {author} {\bibfnamefont {C.}~\bibnamefont {Smith}}, \bibinfo {author} {\bibfnamefont {Y.}~\bibnamefont {Chen}}, \bibinfo {author} {\bibfnamefont {R.}~\bibnamefont {Levy}}, \bibinfo {author} {\bibfnamefont {Y.}~\bibnamefont {Yang}}, \bibinfo {author} {\bibfnamefont {M.~A.}\ \bibnamefont {Morales}},\ and\ \bibinfo {author} {\bibfnamefont {S.}~\bibnamefont {Zhang}},\ }\bibfield  {title} {\bibinfo {title} {Unified variational approach description of ground-state phases of the two-dimensional electron gas},\ }\href {https://doi.org/10.1103/PhysRevLett.133.266504} {\bibfield  {journal} {\bibinfo  {journal} {Phys. Rev. Lett.}\ }\textbf {\bibinfo {volume} {133}},\ \bibinfo {pages} {266504} (\bibinfo {year} {2024})}\BibitemShut {NoStop}%
\bibitem [{\citenamefont {Tsui}\ \emph {et~al.}(2024)\citenamefont {Tsui}, \citenamefont {He}, \citenamefont {Hu}, \citenamefont {Lake}, \citenamefont {Wang}, \citenamefont {Watanabe}, \citenamefont {Taniguchi}, \citenamefont {Zaletel},\ and\ \citenamefont {Yazdani}}]{Tsui2024}%
  \BibitemOpen
  \bibfield  {author} {\bibinfo {author} {\bibfnamefont {Y.-C.}\ \bibnamefont {Tsui}}, \bibinfo {author} {\bibfnamefont {M.}~\bibnamefont {He}}, \bibinfo {author} {\bibfnamefont {Y.}~\bibnamefont {Hu}}, \bibinfo {author} {\bibfnamefont {E.}~\bibnamefont {Lake}}, \bibinfo {author} {\bibfnamefont {T.}~\bibnamefont {Wang}}, \bibinfo {author} {\bibfnamefont {K.}~\bibnamefont {Watanabe}}, \bibinfo {author} {\bibfnamefont {T.}~\bibnamefont {Taniguchi}}, \bibinfo {author} {\bibfnamefont {M.~P.}\ \bibnamefont {Zaletel}},\ and\ \bibinfo {author} {\bibfnamefont {A.}~\bibnamefont {Yazdani}},\ }\bibfield  {title} {\bibinfo {title} {Direct observation of a magnetic-field-induced wigner crystal},\ }\href {https://doi.org/10.1038/s41586-024-07212-7} {\bibfield  {journal} {\bibinfo  {journal} {Nature}\ }\textbf {\bibinfo {volume} {628}},\ \bibinfo {pages} {287} (\bibinfo {year} {2024})}\BibitemShut {NoStop}%
\bibitem [{\citenamefont {Xu}\ \emph {et~al.}(2020)\citenamefont {Xu}, \citenamefont {Liu}, \citenamefont {Rhodes}, \citenamefont {Watanabe}, \citenamefont {Taniguchi}, \citenamefont {Hone}, \citenamefont {Elser}, \citenamefont {Mak},\ and\ \citenamefont {Shan}}]{mak_xu2020correlated}%
  \BibitemOpen
  \bibfield  {author} {\bibinfo {author} {\bibfnamefont {Y.}~\bibnamefont {Xu}}, \bibinfo {author} {\bibfnamefont {S.}~\bibnamefont {Liu}}, \bibinfo {author} {\bibfnamefont {D.~A.}\ \bibnamefont {Rhodes}}, \bibinfo {author} {\bibfnamefont {K.}~\bibnamefont {Watanabe}}, \bibinfo {author} {\bibfnamefont {T.}~\bibnamefont {Taniguchi}}, \bibinfo {author} {\bibfnamefont {J.}~\bibnamefont {Hone}}, \bibinfo {author} {\bibfnamefont {V.}~\bibnamefont {Elser}}, \bibinfo {author} {\bibfnamefont {K.~F.}\ \bibnamefont {Mak}},\ and\ \bibinfo {author} {\bibfnamefont {J.}~\bibnamefont {Shan}},\ }\bibfield  {title} {\bibinfo {title} {Correlated insulating states at fractional fillings of moiré superlattices},\ }\href {https://doi.org/10.1038/s41586-020-2868-6} {\bibfield  {journal} {\bibinfo  {journal} {Nature}\ }\textbf {\bibinfo {volume} {587}},\ \bibinfo {pages} {214–218} (\bibinfo {year} {2020})}\BibitemShut {NoStop}%
\bibitem [{\citenamefont {Li}\ \emph {et~al.}(2021)\citenamefont {Li}, \citenamefont {Li}, \citenamefont {Regan}, \citenamefont {Wang}, \citenamefont {Zhao}, \citenamefont {Kahn}, \citenamefont {Yumigeta}, \citenamefont {Blei}, \citenamefont {Taniguchi}, \citenamefont {Watanabe}, \citenamefont {Tongay}, \citenamefont {Zettl}, \citenamefont {Crommie},\ and\ \citenamefont {Wang}}]{feng_li2021imaging}%
  \BibitemOpen
  \bibfield  {author} {\bibinfo {author} {\bibfnamefont {H.}~\bibnamefont {Li}}, \bibinfo {author} {\bibfnamefont {S.}~\bibnamefont {Li}}, \bibinfo {author} {\bibfnamefont {E.~C.}\ \bibnamefont {Regan}}, \bibinfo {author} {\bibfnamefont {D.}~\bibnamefont {Wang}}, \bibinfo {author} {\bibfnamefont {W.}~\bibnamefont {Zhao}}, \bibinfo {author} {\bibfnamefont {S.}~\bibnamefont {Kahn}}, \bibinfo {author} {\bibfnamefont {K.}~\bibnamefont {Yumigeta}}, \bibinfo {author} {\bibfnamefont {M.}~\bibnamefont {Blei}}, \bibinfo {author} {\bibfnamefont {T.}~\bibnamefont {Taniguchi}}, \bibinfo {author} {\bibfnamefont {K.}~\bibnamefont {Watanabe}}, \bibinfo {author} {\bibfnamefont {S.}~\bibnamefont {Tongay}}, \bibinfo {author} {\bibfnamefont {A.}~\bibnamefont {Zettl}}, \bibinfo {author} {\bibfnamefont {M.~F.}\ \bibnamefont {Crommie}},\ and\ \bibinfo {author} {\bibfnamefont {F.}~\bibnamefont {Wang}},\ }\bibfield  {title} {\bibinfo {title} {Imaging two-dimensional generalized wigner crystals},\ }\href
  {https://doi.org/10.1038/s41586-021-03874-9} {\bibfield  {journal} {\bibinfo  {journal} {Nature}\ }\textbf {\bibinfo {volume} {597}},\ \bibinfo {pages} {650} (\bibinfo {year} {2021})}\BibitemShut {NoStop}%
\bibitem [{\citenamefont {Jin}\ \emph {et~al.}(2021)\citenamefont {Jin}, \citenamefont {Tao}, \citenamefont {Li}, \citenamefont {Xu}, \citenamefont {Tang}, \citenamefont {Zhu}, \citenamefont {Liu}, \citenamefont {Watanabe}, \citenamefont {Taniguchi}, \citenamefont {Hone}, \citenamefont {Fu}, \citenamefont {Shan},\ and\ \citenamefont {Mak}}]{Jin2021}%
  \BibitemOpen
  \bibfield  {author} {\bibinfo {author} {\bibfnamefont {C.}~\bibnamefont {Jin}}, \bibinfo {author} {\bibfnamefont {Z.}~\bibnamefont {Tao}}, \bibinfo {author} {\bibfnamefont {T.}~\bibnamefont {Li}}, \bibinfo {author} {\bibfnamefont {Y.}~\bibnamefont {Xu}}, \bibinfo {author} {\bibfnamefont {Y.}~\bibnamefont {Tang}}, \bibinfo {author} {\bibfnamefont {J.}~\bibnamefont {Zhu}}, \bibinfo {author} {\bibfnamefont {S.}~\bibnamefont {Liu}}, \bibinfo {author} {\bibfnamefont {K.}~\bibnamefont {Watanabe}}, \bibinfo {author} {\bibfnamefont {T.}~\bibnamefont {Taniguchi}}, \bibinfo {author} {\bibfnamefont {J.~C.}\ \bibnamefont {Hone}}, \bibinfo {author} {\bibfnamefont {L.}~\bibnamefont {Fu}}, \bibinfo {author} {\bibfnamefont {J.}~\bibnamefont {Shan}},\ and\ \bibinfo {author} {\bibfnamefont {K.~F.}\ \bibnamefont {Mak}},\ }\bibfield  {title} {\bibinfo {title} {Stripe phases in wse2/ws2 moiré superlattices},\ }\href {https://doi.org/10.1038/s41563-021-00959-8} {\bibfield  {journal} {\bibinfo  {journal} {Nature Materials}\
  }\textbf {\bibinfo {volume} {20}},\ \bibinfo {pages} {940–944} (\bibinfo {year} {2021})}\BibitemShut {NoStop}%
\bibitem [{\citenamefont {Regan}\ \emph {et~al.}(2020)\citenamefont {Regan}, \citenamefont {Wang}, \citenamefont {Jin}, \citenamefont {Bakti~Utama}, \citenamefont {Gao}, \citenamefont {Wei}, \citenamefont {Zhao}, \citenamefont {Zhao}, \citenamefont {Zhang}, \citenamefont {Yumigeta}, \citenamefont {Blei}, \citenamefont {Carlstr\"{o}m}, \citenamefont {Watanabe}, \citenamefont {Taniguchi}, \citenamefont {Tongay}, \citenamefont {Crommie}, \citenamefont {Zettl},\ and\ \citenamefont {Wang}}]{Regan2020}%
  \BibitemOpen
  \bibfield  {author} {\bibinfo {author} {\bibfnamefont {E.~C.}\ \bibnamefont {Regan}}, \bibinfo {author} {\bibfnamefont {D.}~\bibnamefont {Wang}}, \bibinfo {author} {\bibfnamefont {C.}~\bibnamefont {Jin}}, \bibinfo {author} {\bibfnamefont {M.~I.}\ \bibnamefont {Bakti~Utama}}, \bibinfo {author} {\bibfnamefont {B.}~\bibnamefont {Gao}}, \bibinfo {author} {\bibfnamefont {X.}~\bibnamefont {Wei}}, \bibinfo {author} {\bibfnamefont {S.}~\bibnamefont {Zhao}}, \bibinfo {author} {\bibfnamefont {W.}~\bibnamefont {Zhao}}, \bibinfo {author} {\bibfnamefont {Z.}~\bibnamefont {Zhang}}, \bibinfo {author} {\bibfnamefont {K.}~\bibnamefont {Yumigeta}}, \bibinfo {author} {\bibfnamefont {M.}~\bibnamefont {Blei}}, \bibinfo {author} {\bibfnamefont {J.~D.}\ \bibnamefont {Carlstr\"{o}m}}, \bibinfo {author} {\bibfnamefont {K.}~\bibnamefont {Watanabe}}, \bibinfo {author} {\bibfnamefont {T.}~\bibnamefont {Taniguchi}}, \bibinfo {author} {\bibfnamefont {S.}~\bibnamefont {Tongay}}, \bibinfo {author} {\bibfnamefont {M.}~\bibnamefont
  {Crommie}}, \bibinfo {author} {\bibfnamefont {A.}~\bibnamefont {Zettl}},\ and\ \bibinfo {author} {\bibfnamefont {F.}~\bibnamefont {Wang}},\ }\bibfield  {title} {\bibinfo {title} {Mott and generalized wigner crystal states in wse2/ws2 moiré superlattices},\ }\href {https://doi.org/10.1038/s41586-020-2092-4} {\bibfield  {journal} {\bibinfo  {journal} {Nature}\ }\textbf {\bibinfo {volume} {579}},\ \bibinfo {pages} {359–363} (\bibinfo {year} {2020})}\BibitemShut {NoStop}%
\bibitem [{\citenamefont {Huang}\ \emph {et~al.}(2021)\citenamefont {Huang}, \citenamefont {Wang}, \citenamefont {Miao}, \citenamefont {Wang}, \citenamefont {Li}, \citenamefont {Lian}, \citenamefont {Taniguchi}, \citenamefont {Watanabe}, \citenamefont {Okamoto}, \citenamefont {Xiao}, \citenamefont {Shi},\ and\ \citenamefont {Cui}}]{Huang2021}%
  \BibitemOpen
  \bibfield  {author} {\bibinfo {author} {\bibfnamefont {X.}~\bibnamefont {Huang}}, \bibinfo {author} {\bibfnamefont {T.}~\bibnamefont {Wang}}, \bibinfo {author} {\bibfnamefont {S.}~\bibnamefont {Miao}}, \bibinfo {author} {\bibfnamefont {C.}~\bibnamefont {Wang}}, \bibinfo {author} {\bibfnamefont {Z.}~\bibnamefont {Li}}, \bibinfo {author} {\bibfnamefont {Z.}~\bibnamefont {Lian}}, \bibinfo {author} {\bibfnamefont {T.}~\bibnamefont {Taniguchi}}, \bibinfo {author} {\bibfnamefont {K.}~\bibnamefont {Watanabe}}, \bibinfo {author} {\bibfnamefont {S.}~\bibnamefont {Okamoto}}, \bibinfo {author} {\bibfnamefont {D.}~\bibnamefont {Xiao}}, \bibinfo {author} {\bibfnamefont {S.-F.}\ \bibnamefont {Shi}},\ and\ \bibinfo {author} {\bibfnamefont {Y.-T.}\ \bibnamefont {Cui}},\ }\bibfield  {title} {\bibinfo {title} {Correlated insulating states at fractional fillings of the ws2/wse2 moiré lattice},\ }\href {https://doi.org/10.1038/s41567-021-01171-w} {\bibfield  {journal} {\bibinfo  {journal} {Nature Physics}\ }\textbf {\bibinfo
  {volume} {17}},\ \bibinfo {pages} {715–719} (\bibinfo {year} {2021})}\BibitemShut {NoStop}%
\bibitem [{\citenamefont {Pan}\ \emph {et~al.}(2020)\citenamefont {Pan}, \citenamefont {Wu},\ and\ \citenamefont {Das~Sarma}}]{pan2020quantum}%
  \BibitemOpen
  \bibfield  {author} {\bibinfo {author} {\bibfnamefont {H.}~\bibnamefont {Pan}}, \bibinfo {author} {\bibfnamefont {F.}~\bibnamefont {Wu}},\ and\ \bibinfo {author} {\bibfnamefont {S.}~\bibnamefont {Das~Sarma}},\ }\bibfield  {title} {\bibinfo {title} {Quantum phase diagram of a moir{\'e}-hubbard model},\ }\href@noop {} {\bibfield  {journal} {\bibinfo  {journal} {Physical Review B}\ }\textbf {\bibinfo {volume} {102}},\ \bibinfo {pages} {201104} (\bibinfo {year} {2020})}\BibitemShut {NoStop}%
\bibitem [{\citenamefont {Wu}\ \emph {et~al.}(2018)\citenamefont {Wu}, \citenamefont {Lovorn}, \citenamefont {Tutuc},\ and\ \citenamefont {MacDonald}}]{Wu_Macdonald}%
  \BibitemOpen
  \bibfield  {author} {\bibinfo {author} {\bibfnamefont {F.}~\bibnamefont {Wu}}, \bibinfo {author} {\bibfnamefont {T.}~\bibnamefont {Lovorn}}, \bibinfo {author} {\bibfnamefont {E.}~\bibnamefont {Tutuc}},\ and\ \bibinfo {author} {\bibfnamefont {A.~H.}\ \bibnamefont {MacDonald}},\ }\bibfield  {title} {\bibinfo {title} {Hubbard model physics in transition metal dichalcogenide moir\'e bands},\ }\href {https://doi.org/10.1103/PhysRevLett.121.026402} {\bibfield  {journal} {\bibinfo  {journal} {Phys. Rev. Lett.}\ }\textbf {\bibinfo {volume} {121}},\ \bibinfo {pages} {026402} (\bibinfo {year} {2018})}\BibitemShut {NoStop}%
\bibitem [{\citenamefont {Morales-Dur\'an}\ \emph {et~al.}(2023)\citenamefont {Morales-Dur\'an}, \citenamefont {Potasz},\ and\ \citenamefont {MacDonald}}]{PhysRevB.107.235131}%
  \BibitemOpen
  \bibfield  {author} {\bibinfo {author} {\bibfnamefont {N.}~\bibnamefont {Morales-Dur\'an}}, \bibinfo {author} {\bibfnamefont {P.}~\bibnamefont {Potasz}},\ and\ \bibinfo {author} {\bibfnamefont {A.~H.}\ \bibnamefont {MacDonald}},\ }\bibfield  {title} {\bibinfo {title} {Magnetism and quantum melting in moir\'e-material wigner crystals},\ }\href {https://doi.org/10.1103/PhysRevB.107.235131} {\bibfield  {journal} {\bibinfo  {journal} {Phys. Rev. B}\ }\textbf {\bibinfo {volume} {107}},\ \bibinfo {pages} {235131} (\bibinfo {year} {2023})}\BibitemShut {NoStop}%
\bibitem [{\citenamefont {Ung}\ \emph {et~al.}(2023)\citenamefont {Ung}, \citenamefont {Lee},\ and\ \citenamefont {Reichman}}]{PhysRevB.108.245113}%
  \BibitemOpen
  \bibfield  {author} {\bibinfo {author} {\bibfnamefont {S.~F.}\ \bibnamefont {Ung}}, \bibinfo {author} {\bibfnamefont {J.}~\bibnamefont {Lee}},\ and\ \bibinfo {author} {\bibfnamefont {D.~R.}\ \bibnamefont {Reichman}},\ }\bibfield  {title} {\bibinfo {title} {Competing generalized wigner crystal states in moir\'e heterostructures},\ }\href {https://doi.org/10.1103/PhysRevB.108.245113} {\bibfield  {journal} {\bibinfo  {journal} {Phys. Rev. B}\ }\textbf {\bibinfo {volume} {108}},\ \bibinfo {pages} {245113} (\bibinfo {year} {2023})}\BibitemShut {NoStop}%
\bibitem [{\citenamefont {Zhou}\ \emph {et~al.}(2025)\citenamefont {Zhou}, \citenamefont {Esterlis},\ and\ \citenamefont {Smoleński}}]{Zhou_review_2025}%
  \BibitemOpen
  \bibfield  {author} {\bibinfo {author} {\bibfnamefont {Y.}~\bibnamefont {Zhou}}, \bibinfo {author} {\bibfnamefont {I.}~\bibnamefont {Esterlis}},\ and\ \bibinfo {author} {\bibfnamefont {T.}~\bibnamefont {Smoleński}},\ }\href {https://arxiv.org/abs/2509.21222} {\bibinfo {title} {Electronic crystals in layered materials}} (\bibinfo {year} {2025}),\ \Eprint {https://arxiv.org/abs/2509.21222} {arXiv:2509.21222 [cond-mat.str-el]} \BibitemShut {NoStop}%
\bibitem [{\citenamefont {Motruk}\ \emph {et~al.}(2023)\citenamefont {Motruk}, \citenamefont {Rossi}, \citenamefont {Abanin},\ and\ \citenamefont {Rademaker}}]{Motruk_2023}%
  \BibitemOpen
  \bibfield  {author} {\bibinfo {author} {\bibfnamefont {J.}~\bibnamefont {Motruk}}, \bibinfo {author} {\bibfnamefont {D.}~\bibnamefont {Rossi}}, \bibinfo {author} {\bibfnamefont {D.~A.}\ \bibnamefont {Abanin}},\ and\ \bibinfo {author} {\bibfnamefont {L.}~\bibnamefont {Rademaker}},\ }\bibfield  {title} {\bibinfo {title} {Kagome chiral spin liquid in transition metal dichalcogenide moir\'e bilayers},\ }\href {https://doi.org/10.1103/PhysRevResearch.5.L022049} {\bibfield  {journal} {\bibinfo  {journal} {Phys. Rev. Res.}\ }\textbf {\bibinfo {volume} {5}},\ \bibinfo {pages} {L022049} (\bibinfo {year} {2023})}\BibitemShut {NoStop}%
\bibitem [{\citenamefont {Tan}\ \emph {et~al.}(2023)\citenamefont {Tan}, \citenamefont {Tsang}, \citenamefont {Dobrosavljevi\ifmmode~\acute{c}\else \'{c}\fi{}},\ and\ \citenamefont {Rademaker}}]{Tan_2023}%
  \BibitemOpen
  \bibfield  {author} {\bibinfo {author} {\bibfnamefont {Y.}~\bibnamefont {Tan}}, \bibinfo {author} {\bibfnamefont {P.~K.~H.}\ \bibnamefont {Tsang}}, \bibinfo {author} {\bibfnamefont {V.}~\bibnamefont {Dobrosavljevi\ifmmode~\acute{c}\else \'{c}\fi{}}},\ and\ \bibinfo {author} {\bibfnamefont {L.}~\bibnamefont {Rademaker}},\ }\bibfield  {title} {\bibinfo {title} {Doping a wigner-mott insulator: Exotic charge orders in transition metal dichalcogenide moir\'e heterobilayers},\ }\href {https://doi.org/10.1103/PhysRevResearch.5.043190} {\bibfield  {journal} {\bibinfo  {journal} {Phys. Rev. Res.}\ }\textbf {\bibinfo {volume} {5}},\ \bibinfo {pages} {043190} (\bibinfo {year} {2023})}\BibitemShut {NoStop}%
\bibitem [{\citenamefont {Musser}\ \emph {et~al.}(2022)\citenamefont {Musser}, \citenamefont {Senthil},\ and\ \citenamefont {Chowdhury}}]{PhysRevB.106.155145}%
  \BibitemOpen
  \bibfield  {author} {\bibinfo {author} {\bibfnamefont {S.}~\bibnamefont {Musser}}, \bibinfo {author} {\bibfnamefont {T.}~\bibnamefont {Senthil}},\ and\ \bibinfo {author} {\bibfnamefont {D.}~\bibnamefont {Chowdhury}},\ }\bibfield  {title} {\bibinfo {title} {Theory of a continuous bandwidth-tuned wigner-mott transition},\ }\href {https://doi.org/10.1103/PhysRevB.106.155145} {\bibfield  {journal} {\bibinfo  {journal} {Phys. Rev. B}\ }\textbf {\bibinfo {volume} {106}},\ \bibinfo {pages} {155145} (\bibinfo {year} {2022})}\BibitemShut {NoStop}%
\bibitem [{\citenamefont {Zhang}\ \emph {et~al.}(2021)\citenamefont {Zhang}, \citenamefont {Liu},\ and\ \citenamefont {Fu}}]{PhysRevB.103.155142}%
  \BibitemOpen
  \bibfield  {author} {\bibinfo {author} {\bibfnamefont {Y.}~\bibnamefont {Zhang}}, \bibinfo {author} {\bibfnamefont {T.}~\bibnamefont {Liu}},\ and\ \bibinfo {author} {\bibfnamefont {L.}~\bibnamefont {Fu}},\ }\bibfield  {title} {\bibinfo {title} {Electronic structures, charge transfer, and charge order in twisted transition metal dichalcogenide bilayers},\ }\href {https://doi.org/10.1103/PhysRevB.103.155142} {\bibfield  {journal} {\bibinfo  {journal} {Phys. Rev. B}\ }\textbf {\bibinfo {volume} {103}},\ \bibinfo {pages} {155142} (\bibinfo {year} {2021})}\BibitemShut {NoStop}%
\bibitem [{\citenamefont {Yang}\ \emph {et~al.}(2024)\citenamefont {Yang}, \citenamefont {Morales},\ and\ \citenamefont {Zhang}}]{PhysRevLett.132.076503}%
  \BibitemOpen
  \bibfield  {author} {\bibinfo {author} {\bibfnamefont {Y.}~\bibnamefont {Yang}}, \bibinfo {author} {\bibfnamefont {M.~A.}\ \bibnamefont {Morales}},\ and\ \bibinfo {author} {\bibfnamefont {S.}~\bibnamefont {Zhang}},\ }\bibfield  {title} {\bibinfo {title} {Metal-insulator transition in a semiconductor heterobilayer model},\ }\href {https://doi.org/10.1103/PhysRevLett.132.076503} {\bibfield  {journal} {\bibinfo  {journal} {Phys. Rev. Lett.}\ }\textbf {\bibinfo {volume} {132}},\ \bibinfo {pages} {076503} (\bibinfo {year} {2024})}\BibitemShut {NoStop}%
\bibitem [{\citenamefont {Padhi}\ \emph {et~al.}(2021{\natexlab{a}})\citenamefont {Padhi}, \citenamefont {Chitra},\ and\ \citenamefont {Phillips}}]{PhysRevB.103.125146}%
  \BibitemOpen
  \bibfield  {author} {\bibinfo {author} {\bibfnamefont {B.}~\bibnamefont {Padhi}}, \bibinfo {author} {\bibfnamefont {R.}~\bibnamefont {Chitra}},\ and\ \bibinfo {author} {\bibfnamefont {P.~W.}\ \bibnamefont {Phillips}},\ }\bibfield  {title} {\bibinfo {title} {Generalized wigner crystallization in moir\'e materials},\ }\href {https://doi.org/10.1103/PhysRevB.103.125146} {\bibfield  {journal} {\bibinfo  {journal} {Phys. Rev. B}\ }\textbf {\bibinfo {volume} {103}},\ \bibinfo {pages} {125146} (\bibinfo {year} {2021}{\natexlab{a}})}\BibitemShut {NoStop}%
\bibitem [{\citenamefont {Matty}\ and\ \citenamefont {Kim}(2022)}]{Matty2022}%
  \BibitemOpen
  \bibfield  {author} {\bibinfo {author} {\bibfnamefont {M.}~\bibnamefont {Matty}}\ and\ \bibinfo {author} {\bibfnamefont {E.-A.}\ \bibnamefont {Kim}},\ }\bibfield  {title} {\bibinfo {title} {Melting of generalized wigner crystals in transition metal dichalcogenide heterobilayer moiré systems},\ }\bibfield  {journal} {\bibinfo  {journal} {Nature Communications}\ }\textbf {\bibinfo {volume} {13}},\ \href {https://doi.org/10.1038/s41467-022-34683-x} {10.1038/s41467-022-34683-x} (\bibinfo {year} {2022})\BibitemShut {NoStop}%
\bibitem [{\citenamefont {Morales-Dur\'an}\ \emph {et~al.}(2021)\citenamefont {Morales-Dur\'an}, \citenamefont {MacDonald},\ and\ \citenamefont {Potasz}}]{PhysRevB.103.L241110}%
  \BibitemOpen
  \bibfield  {author} {\bibinfo {author} {\bibfnamefont {N.}~\bibnamefont {Morales-Dur\'an}}, \bibinfo {author} {\bibfnamefont {A.~H.}\ \bibnamefont {MacDonald}},\ and\ \bibinfo {author} {\bibfnamefont {P.}~\bibnamefont {Potasz}},\ }\bibfield  {title} {\bibinfo {title} {Metal-insulator transition in transition metal dichalcogenide heterobilayer moir\'e superlattices},\ }\href {https://doi.org/10.1103/PhysRevB.103.L241110} {\bibfield  {journal} {\bibinfo  {journal} {Phys. Rev. B}\ }\textbf {\bibinfo {volume} {103}},\ \bibinfo {pages} {L241110} (\bibinfo {year} {2021})}\BibitemShut {NoStop}%
\bibitem [{\citenamefont {Padhi}\ \emph {et~al.}(2021{\natexlab{b}})\citenamefont {Padhi}, \citenamefont {Chitra},\ and\ \citenamefont {Phillips}}]{Padhi_2021}%
  \BibitemOpen
  \bibfield  {author} {\bibinfo {author} {\bibfnamefont {B.}~\bibnamefont {Padhi}}, \bibinfo {author} {\bibfnamefont {R.}~\bibnamefont {Chitra}},\ and\ \bibinfo {author} {\bibfnamefont {P.~W.}\ \bibnamefont {Phillips}},\ }\bibfield  {title} {\bibinfo {title} {Generalized wigner crystallization in moir\'e materials},\ }\href {https://doi.org/10.1103/PhysRevB.103.125146} {\bibfield  {journal} {\bibinfo  {journal} {Phys. Rev. B}\ }\textbf {\bibinfo {volume} {103}},\ \bibinfo {pages} {125146} (\bibinfo {year} {2021}{\natexlab{b}})}\BibitemShut {NoStop}%
\bibitem [{\citenamefont {Biborski}\ and\ \citenamefont {Zegrodnik}(2024)}]{arxiv.2409.11202}%
  \BibitemOpen
  \bibfield  {author} {\bibinfo {author} {\bibfnamefont {A.}~\bibnamefont {Biborski}}\ and\ \bibinfo {author} {\bibfnamefont {M.}~\bibnamefont {Zegrodnik}},\ }\bibfield  {title} {\bibinfo {title} {Charge and spin properties of a generalized wigner crystal realized in the moiré wse$_2$/ws$_2$ heterobilayer},\ }\href@noop {} {\bibfield  {journal} {\bibinfo  {journal} {arXiv preprint arXiv:2409.11202}\ } (\bibinfo {year} {2024})}\BibitemShut {NoStop}%
\bibitem [{\citenamefont {Kumar}\ \emph {et~al.}(2025)\citenamefont {Kumar}, \citenamefont {Lewandowski},\ and\ \citenamefont {Changlani}}]{kumar_wigner_2025}%
  \BibitemOpen
  \bibfield  {author} {\bibinfo {author} {\bibfnamefont {A.}~\bibnamefont {Kumar}}, \bibinfo {author} {\bibfnamefont {C.}~\bibnamefont {Lewandowski}},\ and\ \bibinfo {author} {\bibfnamefont {H.~J.}\ \bibnamefont {Changlani}},\ }\bibfield  {title} {\bibinfo {title} {\textit{Origin and stability of generalized Wigner crystallinity in triangular moir{\'e} systems}},\ }\href {https://doi.org/10.1038/s41535-025-00792-1} {\bibfield  {journal} {\bibinfo  {journal} {npj Quantum Materials}\ }\textbf {\bibinfo {volume} {10}},\ \bibinfo {pages} {95} (\bibinfo {year} {2025})}\BibitemShut {NoStop}%
\bibitem [{\citenamefont {Tang}\ \emph {et~al.}(2022)\citenamefont {Tang}, \citenamefont {Gu}, \citenamefont {Liu}, \citenamefont {Watanabe}, \citenamefont {Taniguchi}, \citenamefont {Hone}, \citenamefont {Mak},\ and\ \citenamefont {Shan}}]{tang_natcomm_2022}%
  \BibitemOpen
  \bibfield  {author} {\bibinfo {author} {\bibfnamefont {Y.}~\bibnamefont {Tang}}, \bibinfo {author} {\bibfnamefont {J.}~\bibnamefont {Gu}}, \bibinfo {author} {\bibfnamefont {S.}~\bibnamefont {Liu}}, \bibinfo {author} {\bibfnamefont {K.}~\bibnamefont {Watanabe}}, \bibinfo {author} {\bibfnamefont {T.}~\bibnamefont {Taniguchi}}, \bibinfo {author} {\bibfnamefont {J.~C.}\ \bibnamefont {Hone}}, \bibinfo {author} {\bibfnamefont {K.~F.}\ \bibnamefont {Mak}},\ and\ \bibinfo {author} {\bibfnamefont {J.}~\bibnamefont {Shan}},\ }\bibfield  {title} {\bibinfo {title} {Dielectric catastrophe at the wigner-mott transition in a moir{\'e} superlattice},\ }\href {https://doi.org/10.1038/s41467-022-32037-1} {\bibfield  {journal} {\bibinfo  {journal} {Nature Communications}\ }\textbf {\bibinfo {volume} {13}},\ \bibinfo {pages} {4271} (\bibinfo {year} {2022})}\BibitemShut {NoStop}%
\bibitem [{\citenamefont {Zhou}\ \emph {et~al.}(2024{\natexlab{a}})\citenamefont {Zhou}, \citenamefont {Sheng},\ and\ \citenamefont {Kim}}]{zhou2023quantum}%
  \BibitemOpen
  \bibfield  {author} {\bibinfo {author} {\bibfnamefont {Y.}~\bibnamefont {Zhou}}, \bibinfo {author} {\bibfnamefont {D.~N.}\ \bibnamefont {Sheng}},\ and\ \bibinfo {author} {\bibfnamefont {E.-A.}\ \bibnamefont {Kim}},\ }\bibfield  {title} {\bibinfo {title} {Quantum melting of generalized wigner crystals in transition metal dichalcogenide moir\'e systems},\ }\href {https://doi.org/10.1103/PhysRevLett.133.156501} {\bibfield  {journal} {\bibinfo  {journal} {Phys. Rev. Lett.}\ }\textbf {\bibinfo {volume} {133}},\ \bibinfo {pages} {156501} (\bibinfo {year} {2024}{\natexlab{a}})}\BibitemShut {NoStop}%
\bibitem [{\citenamefont {Jakli\ifmmode~\check{c}\else \v{c}\fi{}}\ and\ \citenamefont {Prelov\ifmmode~\check{s}\else \v{s}\fi{}ek}(1994)}]{Jaklic_Prelovsek}%
  \BibitemOpen
  \bibfield  {author} {\bibinfo {author} {\bibfnamefont {J.}~\bibnamefont {Jakli\ifmmode~\check{c}\else \v{c}\fi{}}}\ and\ \bibinfo {author} {\bibfnamefont {P.}~\bibnamefont {Prelov\ifmmode~\check{s}\else \v{s}\fi{}ek}},\ }\bibfield  {title} {\bibinfo {title} {Lanczos method for the calculation of finite-temperature quantities in correlated systems},\ }\href {https://doi.org/10.1103/PhysRevB.49.5065} {\bibfield  {journal} {\bibinfo  {journal} {Phys. Rev. B}\ }\textbf {\bibinfo {volume} {49}},\ \bibinfo {pages} {5065} (\bibinfo {year} {1994})}\BibitemShut {NoStop}%
\bibitem [{\citenamefont {Prelov{\v{s}}ek}\ and\ \citenamefont {Bon{\v{c}}a}(2013)}]{Prelovsek_2013}%
  \BibitemOpen
  \bibfield  {author} {\bibinfo {author} {\bibfnamefont {P.}~\bibnamefont {Prelov{\v{s}}ek}}\ and\ \bibinfo {author} {\bibfnamefont {J.}~\bibnamefont {Bon{\v{c}}a}},\ }\bibinfo {title} {Ground state and finite temperature lanczos methods},\ in\ \href {https://doi.org/10.1007/978-3-642-35106-8_1} {\emph {\bibinfo {booktitle} {Strongly Correlated Systems: Numerical Methods}}},\ \bibinfo {editor} {edited by\ \bibinfo {editor} {\bibfnamefont {A.}~\bibnamefont {Avella}}\ and\ \bibinfo {editor} {\bibfnamefont {F.}~\bibnamefont {Mancini}}}\ (\bibinfo  {publisher} {Springer Berlin Heidelberg},\ \bibinfo {address} {Berlin, Heidelberg},\ \bibinfo {year} {2013})\ pp.\ \bibinfo {pages} {1--30}\BibitemShut {NoStop}%
\bibitem [{\citenamefont {Lee}\ \emph {et~al.}(2023)\citenamefont {Lee}, \citenamefont {Sharma}, \citenamefont {Vafek},\ and\ \citenamefont {Changlani}}]{lee2023triangular}%
  \BibitemOpen
  \bibfield  {author} {\bibinfo {author} {\bibfnamefont {K.}~\bibnamefont {Lee}}, \bibinfo {author} {\bibfnamefont {P.}~\bibnamefont {Sharma}}, \bibinfo {author} {\bibfnamefont {O.}~\bibnamefont {Vafek}},\ and\ \bibinfo {author} {\bibfnamefont {H.~J.}\ \bibnamefont {Changlani}},\ }\bibfield  {title} {\bibinfo {title} {\textit{Triangular lattice Hubbard model physics at intermediate temperatures}},\ }\href {https://doi.org/10.1103/PhysRevB.107.235105} {\bibfield  {journal} {\bibinfo  {journal} {Physical Review B}\ }\textbf {\bibinfo {volume} {107}},\ \bibinfo {pages} {235105} (\bibinfo {year} {2023})}\BibitemShut {NoStop}%
\bibitem [{Note1()}]{Note1}%
  \BibitemOpen
  \bibinfo {note} {A more detailed analysis of the physics associated with the ground state phase transitions, with spin incorporated, will be addressed elsewhere.}\BibitemShut {Stop}%
\bibitem [{\citenamefont {Sch\"uler}\ \emph {et~al.}(2013)\citenamefont {Sch\"uler}, \citenamefont {R\"osner}, \citenamefont {Wehling}, \citenamefont {Lichtenstein},\ and\ \citenamefont {Katsnelson}}]{Schuler2013}%
  \BibitemOpen
  \bibfield  {author} {\bibinfo {author} {\bibfnamefont {M.}~\bibnamefont {Sch\"uler}}, \bibinfo {author} {\bibfnamefont {M.}~\bibnamefont {R\"osner}}, \bibinfo {author} {\bibfnamefont {T.~O.}\ \bibnamefont {Wehling}}, \bibinfo {author} {\bibfnamefont {A.~I.}\ \bibnamefont {Lichtenstein}},\ and\ \bibinfo {author} {\bibfnamefont {M.~I.}\ \bibnamefont {Katsnelson}},\ }\bibfield  {title} {\bibinfo {title} {Optimal {Hubbard} {Models} for {Materials} with {Nonlocal} {Coulomb} {Interactions}: {Graphene}, {Silicene}, and {Benzene}},\ }\href@noop {} {\bibfield  {journal} {\bibinfo  {journal} {Physical Review Letters}\ }\textbf {\bibinfo {volume} {111}},\ \bibinfo {pages} {036601} (\bibinfo {year} {2013})}\BibitemShut {NoStop}%
\bibitem [{\citenamefont {Changlani}\ \emph {et~al.}(2015)\citenamefont {Changlani}, \citenamefont {Zheng},\ and\ \citenamefont {Wagner}}]{Changlani_downfolding_2015}%
  \BibitemOpen
  \bibfield  {author} {\bibinfo {author} {\bibfnamefont {H.~J.}\ \bibnamefont {Changlani}}, \bibinfo {author} {\bibfnamefont {H.}~\bibnamefont {Zheng}},\ and\ \bibinfo {author} {\bibfnamefont {L.~K.}\ \bibnamefont {Wagner}},\ }\bibfield  {title} {\bibinfo {title} {{Density-matrix based determination of low-energy model Hamiltonians from ab initio wavefunctions}},\ }\href {https://doi.org/10.1063/1.4927664} {\bibfield  {journal} {\bibinfo  {journal} {The Journal of Chemical Physics}\ }\textbf {\bibinfo {volume} {143}},\ \bibinfo {pages} {102814} (\bibinfo {year} {2015})},\ \Eprint {https://arxiv.org/abs/https://pubs.aip.org/aip/jcp/article-pdf/doi/10.1063/1.4927664/15502754/102814\_1\_online.pdf} {https://pubs.aip.org/aip/jcp/article-pdf/doi/10.1063/1.4927664/15502754/102814\_1\_online.pdf} \BibitemShut {NoStop}%
\bibitem [{\citenamefont {Zheng}\ \emph {et~al.}(2018)\citenamefont {Zheng}, \citenamefont {Changlani}, \citenamefont {Williams}, \citenamefont {Busemeyer},\ and\ \citenamefont {Wagner}}]{Zheng_downfolding_2018}%
  \BibitemOpen
  \bibfield  {author} {\bibinfo {author} {\bibfnamefont {H.}~\bibnamefont {Zheng}}, \bibinfo {author} {\bibfnamefont {H.~J.}\ \bibnamefont {Changlani}}, \bibinfo {author} {\bibfnamefont {K.~T.}\ \bibnamefont {Williams}}, \bibinfo {author} {\bibfnamefont {B.}~\bibnamefont {Busemeyer}},\ and\ \bibinfo {author} {\bibfnamefont {L.~K.}\ \bibnamefont {Wagner}},\ }\bibfield  {title} {\bibinfo {title} {From real materials to model hamiltonians with density matrix downfolding},\ }\bibfield  {journal} {\bibinfo  {journal} {Frontiers in Physics}\ }\textbf {\bibinfo {volume} {6}},\ \href {https://doi.org/10.3389/fphy.2018.00043} {10.3389/fphy.2018.00043} (\bibinfo {year} {2018})\BibitemShut {NoStop}%
\bibitem [{\citenamefont {Zhou}\ \emph {et~al.}(2024{\natexlab{b}})\citenamefont {Zhou}, \citenamefont {Sheng},\ and\ \citenamefont {Kim}}]{Zhou_PRL_2024}%
  \BibitemOpen
  \bibfield  {author} {\bibinfo {author} {\bibfnamefont {Y.}~\bibnamefont {Zhou}}, \bibinfo {author} {\bibfnamefont {D.~N.}\ \bibnamefont {Sheng}},\ and\ \bibinfo {author} {\bibfnamefont {E.-A.}\ \bibnamefont {Kim}},\ }\bibfield  {title} {\bibinfo {title} {Quantum melting of generalized wigner crystals in transition metal dichalcogenide moir\'e systems},\ }\href {https://doi.org/10.1103/PhysRevLett.133.156501} {\bibfield  {journal} {\bibinfo  {journal} {Phys. Rev. Lett.}\ }\textbf {\bibinfo {volume} {133}},\ \bibinfo {pages} {156501} (\bibinfo {year} {2024}{\natexlab{b}})}\BibitemShut {NoStop}%
\bibitem [{\citenamefont {Hotta}\ and\ \citenamefont {Furukawa}(2006)}]{Hotta_2006}%
  \BibitemOpen
  \bibfield  {author} {\bibinfo {author} {\bibfnamefont {C.}~\bibnamefont {Hotta}}\ and\ \bibinfo {author} {\bibfnamefont {N.}~\bibnamefont {Furukawa}},\ }\bibfield  {title} {\bibinfo {title} {Strong coupling theory of the spinless charges on triangular lattices: Possible formation of a gapless charge-ordered liquid},\ }\href {https://doi.org/10.1103/PhysRevB.74.193107} {\bibfield  {journal} {\bibinfo  {journal} {Phys. Rev. B}\ }\textbf {\bibinfo {volume} {74}},\ \bibinfo {pages} {193107} (\bibinfo {year} {2006})}\BibitemShut {NoStop}%
\bibitem [{\citenamefont {Tocchio}\ \emph {et~al.}(2014)\citenamefont {Tocchio}, \citenamefont {Gros}, \citenamefont {Zhang},\ and\ \citenamefont {Eggert}}]{Tocchio_PRL2014}%
  \BibitemOpen
  \bibfield  {author} {\bibinfo {author} {\bibfnamefont {L.~F.}\ \bibnamefont {Tocchio}}, \bibinfo {author} {\bibfnamefont {C.}~\bibnamefont {Gros}}, \bibinfo {author} {\bibfnamefont {X.-F.}\ \bibnamefont {Zhang}},\ and\ \bibinfo {author} {\bibfnamefont {S.}~\bibnamefont {Eggert}},\ }\bibfield  {title} {\bibinfo {title} {Phase diagram of the triangular extended hubbard model},\ }\href {https://doi.org/10.1103/PhysRevLett.113.246405} {\bibfield  {journal} {\bibinfo  {journal} {Phys. Rev. Lett.}\ }\textbf {\bibinfo {volume} {113}},\ \bibinfo {pages} {246405} (\bibinfo {year} {2014})}\BibitemShut {NoStop}%
\bibitem [{\citenamefont {Merino}\ \emph {et~al.}(2013)\citenamefont {Merino}, \citenamefont {Ralko},\ and\ \citenamefont {Fratini}}]{Fratini_pinball}%
  \BibitemOpen
  \bibfield  {author} {\bibinfo {author} {\bibfnamefont {J.}~\bibnamefont {Merino}}, \bibinfo {author} {\bibfnamefont {A.}~\bibnamefont {Ralko}},\ and\ \bibinfo {author} {\bibfnamefont {S.}~\bibnamefont {Fratini}},\ }\bibfield  {title} {\bibinfo {title} {Emergent heavy fermion behavior at the wigner-mott transition},\ }\href {https://doi.org/10.1103/PhysRevLett.111.126403} {\bibfield  {journal} {\bibinfo  {journal} {Phys. Rev. Lett.}\ }\textbf {\bibinfo {volume} {111}},\ \bibinfo {pages} {126403} (\bibinfo {year} {2013})}\BibitemShut {NoStop}%
\bibitem [{Note2()}]{Note2}%
  \BibitemOpen
  \bibinfo {note} {A more detailed study of the $n=1/2$ NN case will be presented elsewhere.}\BibitemShut {Stop}%
\bibitem [{\citenamefont {Villain}\ \emph {et~al.}(1980)\citenamefont {Villain}, \citenamefont {Bidaux}, \citenamefont {Carton},\ and\ \citenamefont {Conte}}]{Villain_1980}%
  \BibitemOpen
  \bibfield  {author} {\bibinfo {author} {\bibfnamefont {J.}~\bibnamefont {Villain}}, \bibinfo {author} {\bibfnamefont {R.}~\bibnamefont {Bidaux}}, \bibinfo {author} {\bibfnamefont {J.-P.}\ \bibnamefont {Carton}},\ and\ \bibinfo {author} {\bibfnamefont {R.}~\bibnamefont {Conte}},\ }\bibfield  {title} {\bibinfo {title} {{Order as an effect of disorder}},\ }\href {https://doi.org/10.1051/jphys:0198000410110126300} {\bibfield  {journal} {\bibinfo  {journal} {{Journal de Physique}}\ }\textbf {\bibinfo {volume} {41}},\ \bibinfo {pages} {1263} (\bibinfo {year} {1980})}\BibitemShut {NoStop}%
\bibitem [{\citenamefont {Henley}(1989)}]{Henley_obd_1989}%
  \BibitemOpen
  \bibfield  {author} {\bibinfo {author} {\bibfnamefont {C.~L.}\ \bibnamefont {Henley}},\ }\bibfield  {title} {\bibinfo {title} {Ordering due to disorder in a frustrated vector antiferromagnet},\ }\href {https://doi.org/10.1103/PhysRevLett.62.2056} {\bibfield  {journal} {\bibinfo  {journal} {Phys. Rev. Lett.}\ }\textbf {\bibinfo {volume} {62}},\ \bibinfo {pages} {2056} (\bibinfo {year} {1989})}\BibitemShut {NoStop}%
\bibitem [{\citenamefont {Hammam}\ \emph {et~al.}(2025)\citenamefont {Hammam}, \citenamefont {Lewandowski}, \citenamefont {Dobrosavljevic},\ and\ \citenamefont {Joy}}]{hammam2025}%
  \BibitemOpen
  \bibfield  {author} {\bibinfo {author} {\bibfnamefont {M.}~\bibnamefont {Hammam}}, \bibinfo {author} {\bibfnamefont {C.}~\bibnamefont {Lewandowski}}, \bibinfo {author} {\bibfnamefont {V.}~\bibnamefont {Dobrosavljevic}},\ and\ \bibinfo {author} {\bibfnamefont {S.}~\bibnamefont {Joy}},\ }\href {https://arxiv.org/abs/2512.07932} {\bibinfo {title} {Is disorder a friend or a foe to melting of wigner-mott insulators?}} (\bibinfo {year} {2025}),\ \Eprint {https://arxiv.org/abs/2512.07932} {arXiv:2512.07932 [cond-mat.str-el]} \BibitemShut {NoStop}%
\bibitem [{\citenamefont {{Jain}}\ and\ \citenamefont {{Huang}}(2025)}]{Jain_2025}%
  \BibitemOpen
  \bibfield  {author} {\bibinfo {author} {\bibfnamefont {A.}~\bibnamefont {{Jain}}}\ and\ \bibinfo {author} {\bibfnamefont {C.}~\bibnamefont {{Huang}}},\ }\bibfield  {title} {\bibinfo {title} {{Elementary Excitations, Melting Temperature and Correlation Energy in Wigner Crystal}},\ }\href {https://doi.org/10.48550/arXiv.2504.04538} {\bibfield  {journal} {\bibinfo  {journal} {arXiv e-prints}\ ,\ \bibinfo {eid} {arXiv:2504.04538}} (\bibinfo {year} {2025})},\ \Eprint {https://arxiv.org/abs/2504.04538} {arXiv:2504.04538 [cond-mat.str-el]} \BibitemShut {NoStop}%
\bibitem [{\citenamefont {{Changlani}}(2017)}]{Changlani_YbTO}%
  \BibitemOpen
  \bibfield  {author} {\bibinfo {author} {\bibfnamefont {H.~J.}\ \bibnamefont {{Changlani}}},\ }\bibfield  {title} {\bibinfo {title} {{Quantum versus classical effects at zero and finite temperature in the quantum pyrochlore Yb$_2$Ti$_2$O$_7$}},\ }\href@noop {} {\bibfield  {journal} {\bibinfo  {journal} {arXiv e-prints}\ ,\ \bibinfo {eid} {arXiv:1710.02234}} (\bibinfo {year} {2017})}\BibitemShut {NoStop}%
\bibitem [{\citenamefont {Scheie}\ \emph {et~al.}(2017)\citenamefont {Scheie}, \citenamefont {Kindervater}, \citenamefont {S\"aubert}, \citenamefont {Duvinage}, \citenamefont {Pfleiderer}, \citenamefont {Changlani}, \citenamefont {Zhang}, \citenamefont {Harriger}, \citenamefont {Arpino}, \citenamefont {Koohpayeh}, \citenamefont {Tchernyshyov},\ and\ \citenamefont {Broholm}}]{Scheie2017}%
  \BibitemOpen
  \bibfield  {author} {\bibinfo {author} {\bibfnamefont {A.}~\bibnamefont {Scheie}}, \bibinfo {author} {\bibfnamefont {J.}~\bibnamefont {Kindervater}}, \bibinfo {author} {\bibfnamefont {S.}~\bibnamefont {S\"aubert}}, \bibinfo {author} {\bibfnamefont {C.}~\bibnamefont {Duvinage}}, \bibinfo {author} {\bibfnamefont {C.}~\bibnamefont {Pfleiderer}}, \bibinfo {author} {\bibfnamefont {H.~J.}\ \bibnamefont {Changlani}}, \bibinfo {author} {\bibfnamefont {S.}~\bibnamefont {Zhang}}, \bibinfo {author} {\bibfnamefont {L.}~\bibnamefont {Harriger}}, \bibinfo {author} {\bibfnamefont {K.}~\bibnamefont {Arpino}}, \bibinfo {author} {\bibfnamefont {S.~M.}\ \bibnamefont {Koohpayeh}}, \bibinfo {author} {\bibfnamefont {O.}~\bibnamefont {Tchernyshyov}},\ and\ \bibinfo {author} {\bibfnamefont {C.}~\bibnamefont {Broholm}},\ }\bibfield  {title} {\bibinfo {title} {Reentrant phase diagram of {$\rm Yb_2Ti_2O_7$} in a $\langle 111 \rangle$ magnetic field},\ }\href {https://doi.org/10.1103/PhysRevLett.119.127201} {\bibfield  {journal}
  {\bibinfo  {journal} {Phys. Rev. Lett.}\ }\textbf {\bibinfo {volume} {119}},\ \bibinfo {pages} {127201} (\bibinfo {year} {2017})}\BibitemShut {NoStop}%
\bibitem [{\citenamefont {Hallas}\ \emph {et~al.}(2018)\citenamefont {Hallas}, \citenamefont {Gaudet},\ and\ \citenamefont {Gaulin}}]{hallas2018experimental}%
  \BibitemOpen
  \bibfield  {author} {\bibinfo {author} {\bibfnamefont {A.~M.}\ \bibnamefont {Hallas}}, \bibinfo {author} {\bibfnamefont {J.}~\bibnamefont {Gaudet}},\ and\ \bibinfo {author} {\bibfnamefont {B.~D.}\ \bibnamefont {Gaulin}},\ }\bibfield  {title} {\bibinfo {title} {Experimental insights into ground-state selection of quantum xy pyrochlores},\ }\href {https://doi.org/https://doi.org/10.1146/annurev-conmatphys-031016-025218} {\bibfield  {journal} {\bibinfo  {journal} {Annual Review of Condensed Matter Physics}\ }\textbf {\bibinfo {volume} {9}},\ \bibinfo {pages} {105} (\bibinfo {year} {2018})}\BibitemShut {NoStop}%
\bibitem [{\citenamefont {Melko}\ \emph {et~al.}(2005)\citenamefont {Melko}, \citenamefont {Paramekanti}, \citenamefont {Burkov}, \citenamefont {Vishwanath}, \citenamefont {Sheng},\ and\ \citenamefont {Balents}}]{Melko_PRL_2005}%
  \BibitemOpen
  \bibfield  {author} {\bibinfo {author} {\bibfnamefont {R.~G.}\ \bibnamefont {Melko}}, \bibinfo {author} {\bibfnamefont {A.}~\bibnamefont {Paramekanti}}, \bibinfo {author} {\bibfnamefont {A.~A.}\ \bibnamefont {Burkov}}, \bibinfo {author} {\bibfnamefont {A.}~\bibnamefont {Vishwanath}}, \bibinfo {author} {\bibfnamefont {D.~N.}\ \bibnamefont {Sheng}},\ and\ \bibinfo {author} {\bibfnamefont {L.}~\bibnamefont {Balents}},\ }\bibfield  {title} {\bibinfo {title} {Supersolid order from disorder: Hard-core bosons on the triangular lattice},\ }\href {https://doi.org/10.1103/PhysRevLett.95.127207} {\bibfield  {journal} {\bibinfo  {journal} {Phys. Rev. Lett.}\ }\textbf {\bibinfo {volume} {95}},\ \bibinfo {pages} {127207} (\bibinfo {year} {2005})}\BibitemShut {NoStop}%
\bibitem [{\citenamefont {Boninsegni}\ and\ \citenamefont {Prokof'ev}(2005)}]{Boninsegni_PRL_2005}%
  \BibitemOpen
  \bibfield  {author} {\bibinfo {author} {\bibfnamefont {M.}~\bibnamefont {Boninsegni}}\ and\ \bibinfo {author} {\bibfnamefont {N.}~\bibnamefont {Prokof'ev}},\ }\bibfield  {title} {\bibinfo {title} {Supersolid phase of hard-core bosons on a triangular lattice},\ }\href {https://doi.org/10.1103/PhysRevLett.95.237204} {\bibfield  {journal} {\bibinfo  {journal} {Phys. Rev. Lett.}\ }\textbf {\bibinfo {volume} {95}},\ \bibinfo {pages} {237204} (\bibinfo {year} {2005})}\BibitemShut {NoStop}%
\bibitem [{\citenamefont {Heidarian}\ and\ \citenamefont {Damle}(2005)}]{Heidarian_PRL_2005}%
  \BibitemOpen
  \bibfield  {author} {\bibinfo {author} {\bibfnamefont {D.}~\bibnamefont {Heidarian}}\ and\ \bibinfo {author} {\bibfnamefont {K.}~\bibnamefont {Damle}},\ }\bibfield  {title} {\bibinfo {title} {Persistent supersolid phase of hard-core bosons on the triangular lattice},\ }\href {https://doi.org/10.1103/PhysRevLett.95.127206} {\bibfield  {journal} {\bibinfo  {journal} {Phys. Rev. Lett.}\ }\textbf {\bibinfo {volume} {95}},\ \bibinfo {pages} {127206} (\bibinfo {year} {2005})}\BibitemShut {NoStop}%
\bibitem [{\citenamefont {Weinberg}\ and\ \citenamefont {Bukov}(2017)}]{quspin}%
  \BibitemOpen
  \bibfield  {author} {\bibinfo {author} {\bibfnamefont {P.}~\bibnamefont {Weinberg}}\ and\ \bibinfo {author} {\bibfnamefont {M.}~\bibnamefont {Bukov}},\ }\bibfield  {title} {\bibinfo {title} {\textit{QuSpin: a Python package for dynamics and exact diagonalisation of quantum many body systems part I: spin chains}},\ }\href {https://doi.org/10.21468/SciPostPhys.2.1.003} {\bibfield  {journal} {\bibinfo  {journal} {SciPost Phys.}\ }\textbf {\bibinfo {volume} {2}},\ \bibinfo {pages} {003} (\bibinfo {year} {2017})}\BibitemShut {NoStop}%
\bibitem [{\citenamefont {Marinari}\ and\ \citenamefont {Parisi}(1992)}]{Marinari_1992}%
  \BibitemOpen
  \bibfield  {author} {\bibinfo {author} {\bibfnamefont {E.}~\bibnamefont {Marinari}}\ and\ \bibinfo {author} {\bibfnamefont {G.}~\bibnamefont {Parisi}},\ }\bibfield  {title} {\bibinfo {title} {Simulated tempering: A new monte carlo scheme},\ }\href {https://doi.org/10.1209/0295-5075/19/6/002} {\bibfield  {journal} {\bibinfo  {journal} {Europhysics Letters}\ }\textbf {\bibinfo {volume} {19}},\ \bibinfo {pages} {451} (\bibinfo {year} {1992})}\BibitemShut {NoStop}%
\bibitem [{\citenamefont {Hukushima}\ and\ \citenamefont {Nemoto}(1996)}]{Hukushima_1996}%
  \BibitemOpen
  \bibfield  {author} {\bibinfo {author} {\bibfnamefont {K.}~\bibnamefont {Hukushima}}\ and\ \bibinfo {author} {\bibfnamefont {K.}~\bibnamefont {Nemoto}},\ }\bibfield  {title} {\bibinfo {title} {Exchange monte carlo method and application to spin glass simulations},\ }\href {https://doi.org/10.1143/JPSJ.65.1604} {\bibfield  {journal} {\bibinfo  {journal} {Journal of the Physical Society of Japan}\ }\textbf {\bibinfo {volume} {65}},\ \bibinfo {pages} {1604} (\bibinfo {year} {1996})},\ \Eprint {https://arxiv.org/abs/https://doi.org/10.1143/JPSJ.65.1604} {https://doi.org/10.1143/JPSJ.65.1604} \BibitemShut {NoStop}%
\end{thebibliography}
\end{document}